\documentclass{aa}  

\usepackage{txfonts}

\usepackage{multirow}
\usepackage{array}
\usepackage{newtxtext,newtxmath}
\usepackage{amssymb}
\usepackage{graphicx}
\usepackage{amsmath}
\usepackage{enumitem}
\usepackage{lscape}
\usepackage[T1]{fontenc} 
\usepackage{natbib} 
\usepackage{wrapfig}
\usepackage{float}
\usepackage{siunitx}
\usepackage{dcolumn}

\usepackage{booktabs}

\usepackage{color}
\usepackage{natbib,twoopt}
\usepackage[hyphenbreaks]{breakurl}
\usepackage[breaklinks]{hyperref}      
\bibpunct{(}{)}{;}{a}{}{,}             
\definecolor{cobalt}{rgb}{0.06, 0.2, 0.65}
\hypersetup{
  colorlinks,
  citecolor=cobalt,
  linkcolor=[rgb]{0.8, 0.2, 1.0},
  urlcolor=cobalt,
}
\makeatletter
  \newcommandtwoopt{\citeads}[3][][]{\href{http://adsabs.harvard.edu/abs/#3}%
    {\def\hyper@linkstart##1##2{}%
     \let\hyper@linkend\@empty\citealp[#1][#2]{#3}}}
  \newcommandtwoopt{\citepads}[3][][]{\href{http://adsabs.harvard.edu/abs/#3}%
    {\def\hyper@linkstart##1##2{}%
     \let\hyper@linkend\@empty\citep[#1][#2]{#3}}}
  \newcommandtwoopt{\citetads}[3][][]{\href{http://adsabs.harvard.edu/abs/#3}%
    {\def\hyper@linkstart##1##2{}%
     \let\hyper@linkend\@empty\citet[#1][#2]{#3}}}
  \newcommandtwoopt{\citeyearads}[3][][]%
    {\href{http://adsabs.harvard.edu/abs/#3}
    {\def\hyper@linkstart##1##2{}%
     \let\hyper@linkend\@empty\citeyear[#1][#2]{#3}}}
\makeatother

\begin{document}

\title{Evolution of dust attenuation in star-forming galaxies with UV slope, stellar mass, and redshift out to $z\sim 5$}

\author{J.~V.~Wijesekera\inst{\ref{inst:tor}}
\and
M.~P.~Koprowski\inst{\ref{inst:tor}}
\and
J.~S.~Dunlop\inst{\ref{inst:roe}}
\and
K.~Lisiecki\inst{\ref{inst:tor},\ref{inst:ncbj}}
\and
D.~J.~McLeod\inst{\ref{inst:roe}}
\and
R.~J.~McLure\inst{\ref{inst:roe}}
\and
M.~J.~Micha{\l}owski\inst{\ref{inst:roe},\ref{inst:poz}}
\and
M.~Solar\inst{\ref{inst:poz}}
        }

\institute{
Institute of Astronomy, Faculty of Physics, Astronomy and Informatics, Nicolaus Copernicus University, Grudzi\c{a}dzka 5, 87-100 Toru\'{n}, Poland, {\tt drelkopi@gmail.com}\label{inst:tor}
\and
Institute for Astronomy, University of Edinburgh, Royal Observatory, Edinburgh EH9 3HJ, UK \label{inst:roe}
\and
National Centre for Nuclear Research, Pasteura 7, 093, Warsaw, Poland \label{inst:ncbj}
\and
Astronomical Observatory Institute, Faculty of Physics and Astronomy, Adam Mickiewicz University, ul.~S{\l}oneczna 36, 60-286 Pozna{\'n}, Poland \label{inst:poz}
}

\date{Received September 15, 1996; accepted March 16, 1997}


\abstract
{In order to reliably constrain the dust-obscured portion of star formation across cosmic time, an accurate calibration of the so-called infrared excess (${\rm IRX} \equiv L_{\mathrm{IR}}/L_{\mathrm{UV}}$) and its dependence on intrinsic galaxy properties is required. While the local IRX-$\beta$ relation (where $F_\lambda\propto \lambda^\beta$) has been widely used to correct for the effects of dust absorption and scattering, many recent high-redshift works show significant inconsistencies, being most likely caused by differences in the assumed attenuation curves, dust temperatures and galaxies' star-to-dust relative morphology.}
{We derive a dependence of the IRX on UV slope $\beta$, stellar mass $M_\ast$, and redshift out to $z \simeq 5$, and establish consistent functional relations that can be used for correcting the UV/optical-selected galaxy samples for the effects of dust absorption.}
{This work is based on a $K$-band selected sample of $\sim 10^5$ star-forming galaxies detected in the UDS and COSMOS fields. Quiescent sources and known starbursts are removed, and the IR luminosities are established through stacking in FIR {\it Herschel} and JCMT maps. UV slopes are found from SED fits and stacked IRX values are derived by taking the median of individual IRX measurements in bins of $\beta$, $M_\ast$ and redshift.}
{While our best-fit IRX-$\beta$ relation is consistent with a Calzetti-like attenuation curve at $\beta\gtrsim -1$, at bluer values the IRX seems to increase with redshift due to different mass-completeness limits imposed. When deriving the IRX-$\beta$ relation in stellar-mass bins, a systematic trend is found, where the effective slope of the attenuation law becomes progressively shallower with increasing mass. We incorporate this into the IRX-$\beta$ relation through the slope of the underlying reddening law, $dA_{1600}/d\beta$, being a quadratic function of $\log(M_\ast/{\rm M_\odot})$. Expressing IRX as a function of the stellar mass we find a tight correlation, with IRX rising monotonically with mass but exhibiting a clear high-mass turnover at $z\lesssim 2-3$, consistent with suppressed cold-gas accretion and dust growth in massive systems.}
{}

\keywords{dust,extinction --
            galaxies: evolution --
            galaxies: high-redshift -- galaxies: ISM
           }

\titlerunning{Dust attenuation out to $z\sim 5$}
\authorrunning{Wijesekera et al.}

\maketitle
%

\section{Introduction}
\label{sec:intro}

The accurate determination of the cosmic star-formation rate density ($\rho_{\rm SFR}$) across cosmic time remains a fundamental objective of observational cosmology. To build a complete picture of the star-formation activity in the Universe, one must account for all the energy produced by young, massive stars, which includes both the direct rest-frame ultraviolet (UV) emission that escapes the host galaxy and the fraction absorbed by the interstellar medium (ISM) and re-emitted at far-infrared (FIR) wavelengths. While this task is relatively straightforward in the local Universe due to the wealth of the available high-resolution imaging, at high redshifts ($z > 2$) the impact of the interstellar dust becomes harder to assess (\citealt{Madau_2014} and references therein). 

The standard way to quantify the dust-obscured fraction of the stellar light is through the so-called infrared excess ($\text{IRX} \equiv L_{\rm IR}/L_{\rm UV}$) and its relationship with galaxies' optical properties, like the UV slope ($\beta$) or the stellar mass ($M_\ast$). Historically, the relationship between IRX and $\beta$ has been used as a primary proxy for dust attenuation (e.g., \citealt{Calzetti_2000, Reddy_2012, Bouwens_2016, Fudamoto_2020}). Originally established by \citet{Meurer_1999}, this relationship describes a tight empirical correlation for local starburst galaxies, with a relatively flat, Calzetti-like attenuation law assumed, where the reddening of a galaxy’s UV color between the rest-frame 125-250\,nm can be used to predict its total dust obscuration. For over a decade, this relation was the adopted standard for correcting UV-selected samples at high redshifts. However, with more deep-field surveys producing increasing amount of high-redshift imaging, it became evident that `normal' star-forming galaxies (SFGs) differ significantly from the starburst sequence in terms of the dust content (e.g., \citealt{Reddy_2010, Reddy_2012}). These deviations have been suggested to arise from a combination of varying stellar population ages, metallicity effects, and complex dust-to-star geometries that are not accounted for by the local \citet{Meurer_1999} data (e.g., \citealt{Narayanan_2018, Salim_2020}).

The high-resolution imaging produced by the Atacama Large Millimeter/submillimeter Array (ALMA) has further complicated this picture. Initial FIR observations of Lyman-break galaxies (LBGs) at $z \sim 5-6$ reported surprisingly low IRX values \citep{Capak_2015, Bouwens_2016}, representing a departure from the local Calzetti-like attenuation law typically applied to star-forming systems. This suggested that high-redshift galaxies are potentially more consistent with the Small Magellanic Cloud (SMC)-like extinction law. In addition, wrongly measured or assumed temperature of the interstellar dust ($T_{\rm d}$) was also shown to play a role in the derived values of the IR luminosities. A number of works established $T_{\rm d}$ to be an increasing function of redshift (e.g., \citealt{Viero_2022,Ismail_2023,Koprowski_2024}), with the apparent `IRX deficit' potentially being an artifact of underestimating the FIR luminosity of warm-dust systems.

Furthermore, the IRX–$\beta$ relationship has recently been found to also depend on the stellar masses of the galaxy samples \citep{Alvarez_2019, Bouwens_2020}. It has been found that while low-mass systems at high redshifts appear `IR-faint' for a given value of beta, more massive galaxies, $\log(M_\ast/M_\odot) > 10$, are often consistent with the local Calzetti-like attenuation curves. This mass dependence suggests that as galaxies grow, they rapidly accumulate dust and develop the complex, multi-component ISM structures seen in local starbursts (e.g., \citealt{vanderWel_2025, Gebek_2025, Lisiecki_2025}). Recent JWST observations (e.g., \citealt{Fisher_2025, Shivaei_2025}) further indicate that the scatter in the IRX-$\beta$ plane is amplified by `bursty' star-formation histories, where high UV luminosities can `decouple' from the dust-obscured regions, resulting in dusty sources possessing relatively blue values of the UV slope.

Beyond the complications related to the physical evolution of galaxies, a significant challenge lies in the selection of both rest-frame UV/optical and FIR-based samples. Most high-redshift ALMA studies, for instance, are biased toward the most IR-luminous sources, potentially missing the `normal' population of star-forming galaxies, while samples selected at the rest-frame UV/optical bands are often incomplete in terms of the stellar mass.  To overcome this, we utilize a $K$-band selected sample of approximately 100,000 galaxies from the UKIDSS Ultra Deep Survey (UDS) and COSMOS fields, covering a combined area of nearly 2 deg$^2$ \citep{McLeod_2021}. In order to reach beyond the detection limits of typical FIR observations, we resort to a well-known method of stacking. This approach is critical for ensuring mass completeness and accounting for the selection effects that often plague individual FIR detections studies. By stacking individual infrared excess values of galaxies in distinct UV slope, stellar mass and redshift bins, we were able to establish a consistent relationship between the IRX, UV slope, stellar mass and redshift for star-forming galaxies out to $z\sim 5$. 

This paper is structured as follows. In Section\,\ref{sec:data} we describe the data used in this work. The star-forming galaxies' sample selection process and methods used to derive the physical properties and the stacked IRX values are presented in Section\,\ref{sec:meth}. In Section\,\ref{sec:disc} the results are presented and discussed, where the high-redshift IRX-$\beta$ relation is derived in Section\,\ref{sec:irxb}, the stellar mass dependence through the slope of the reddening law is introduced into the IRX-$\beta$ relationship in Section\,\ref{sec:mirx}, followed by the comparison with recent literature in Section\,\ref{sec:irxbcomp}. The IRX-$M_\ast$ function and its evolution with redshift is presented, discussed and compared with other studies in Sections\,\ref{sec:irxm} and \ref{sec:irxmcomp}. We summarize in Section\,\ref{sec:sum}. Throughout the paper, we use the \citet{Chabrier_2003} stellar IMF with an assumed flat cosmology of $\Omega_{\rm m} = 0.3$, $\Omega_\Lambda = 0.7$, and ${\rm H}_0 = 70$\,km\,s$^{-1}$\,Mpc$^{-1}$.

\section{Data}
\label{sec:data}

\subsection{Optical/near-IR catalogs} \label{sec:opt}

In order to determine how the IRX varies across cosmic time, we stack the optical/near-IR star-forming galaxies' samples of the UKIDSS Ultra Deep Survey (UDS) and COSMOS fields \citep{McLeod_2021} in the FIR {\it Herschel} and James Clerk Maxwell Telescope (JCMT) maps. The details of the data used, the selection process, catalogs construction and the determination of photometric redshifts and stellar masses are presented in \citet{McLeod_2021}, with a short summary given below. 

The UKIDSS UDS field \citep{Lawrence_2007} includes UDS DR11 imaging (Almaini et al., in preparation) in the near-IR JHK bands, Y-band imaging from the VISTA VIDEO DR4 \citep{Jarvis_2013}, UV imaging in $u^\ast$-band from CFHT MegaCam \citep{Gwyn_2012}, optical imaging in BVRiz$^{\prime}$ from Subaru Suprime-Cam \citep{Furusawa_2008} and the mid-infrared \textit{Spitzer} IRAC imaging in 3.6\,$\mu$m and 4.5\,$\mu$m, combining the SEDS \citep{Ashby_2013}, S-CANDELS \citep{Ashby_2015}, and SPLASH (PI Capak; see \citealt{mehta2018}) programs. The effective area of the overlapping UDS imaging was 0.69 deg$^2$, after masking for bright stars. For the COSMOS field, near-IR $YJHK_{\rm s}$ imaging from UltraVISTA DR4 \citep{McCracken_2012} with UV/optical $u^\ast griz$ imaging from the CFHTLS-D2 T0007 data release are combined. As with the UDS, the data also includes the \textit{Spitzer} IRAC imaging available in 3.6\,$\mu$m and 4.5\,$\mu$m, covering the area of 0.86\,deg$^2$.

The optical/near-IR data available for each galaxy were fit with a number of spectral energy distribution (SED) codes, where the final redshift was taken to be the median of individual values, $z_{\rm med}$. The precision was assessed using the available spectroscopic data, with the values of $\sigma_z$, defined as $1.483 \times {\rm MAD}(dz)$, where MAD is the median absolute deviation and $dz = (z_{\rm med} - z_{\rm spec})/(1 + z_{\rm spec})$, for the COSMOS and UDS fields, found to be equal to 0.019 and 0.022, respectively \citep{McLeod_2021}.

Stellar masses were measured for each object by fixing the photometric redshift to $z_{\rm med}$, and re-fitting the SED using {\sc LePhare} with the \citet{Bruzual_2003} library, a \citet{Chabrier_2003} IMF, a \citet{Calzetti_2000} dust attenuation law and IGM absorption as in \citet{Madau_1995}. The resulting values were checked against masses found by the CANDELS team, where a tight 1:1 relation and a typical scatter of $\pm 0.05$ dex was found.

\subsection{Far-infrared} \label{sec:fir}

Two different stacking procedures have been used. For the stacked IRX values in bins of UV slope, stellar mass and redshift, FIR flux densities in the COSMOS and UDS fields were extracted using the SCUBA-2 Cosmology Legacy Survey (S2CLS; \citealt{Geach_2017}) 850\,${\rm \mu m}$ imaging, as explained in Section\,\ref{sec:irx}. In order to determine the functional form of the IRX-$\beta$ relationship, however, our sample was additionally stacked in the {\it Herschel} \citep{Pilbratt_2010} Multi-tiered Extragalactic Survey (HerMES; \citealt{Oliver_2012}) and the Photodetector Array Camera and Spectrometer (PACS; \citealt{Poglitsch_2010}) Evolutionary Probe (PEP; \citealt{Lutz_2011}) maps obtained with the Spectral and Photometric Imaging Receiver (SPIRE; \citealt{Griffin_2010}) and PACS instruments (Section\,\ref{sec:irxb}).

\section{Methods}
\label{sec:meth}

\subsection{Sample selection}
\label{sec:sample}

We binned our data in equal redshift intervals spanning $0.5<z<5.0$ with a step size of $\Delta z=0.5$, with the exception of the highest-redshift bin, where the bin width was set to $\Delta z=1.0$ to improve the detection rate. For the analysis of the IRX-$\beta$ relation, we set the lowest redshift to 2.0, since at lower values the available photometry does not cover the rest-frame UV slope interval of 125-250\,nm. Furthermore, we limited our sample to sources with stellar masses above the 90\% completeness threshold, as determined by \citet{McLeod_2021}.
 
Quiescent galaxies were excluded based on the \textit{NUVrJ} color-color selection criteria of \citet{Ilbert_2013}:

\begin{align}\label{eq:uvj}
\begin{split}
(NUV-r)> &\, 3\times(r-J) + 1;\\
(NUV-r)> &\, 3.1,
\end{split}
\end{align}

\noindent where $NUV$, $r$ and $J$ AB magnitudes are determined from spectral energy distribution (SED) fitting (see Section\,\ref{sec:sed}). This method separates sources whose reddening results from the aging of the stellar population (ie., quiescent galaxies) from the dusty star-forming sample. The \textit{NUVrJ} color-color selection was limited to redshifts less than 4, as the rest-frame $J$-band becomes progressively difficult to trace at higher redshifts with the available photometry.

Similarly to \citet{Koprowski_2024}, we also identified and removed starburst galaxies, using the ALMA-selected samples in the COSMOS \citep{Liu_2019} and UDS \citep{Stach_2019} fields, with stellar masses and SFRs from \citet{Liu_2019} and \citet{Dudzeviciute_2020}, respectively. Following \citet{Elbaz_2018}, starbursts were defined as galaxies with ${\rm SFR/SFR_{MS}}>3$, where ${\rm SFR_{MS}}$ is the corresponding main-sequence value \citep{Koprowski_2024}.

\subsection{SED fitting}
\label{sec:sed}

Spectral energy distribution fitting was performed in order to determine the rest-frame $NUV$, $r$ and $J$ absolute magnitudes used for quiescent galaxies selection (Eq.\,\ref{eq:uvj}), UV slopes, UV luminosities necessary for the derivation of the stacked IRX values and the IRX-$\beta$ relationship functional form parameters (Section\,\ref{sec:irxb}). We used the Code Investigating GALaxy Emission ({\sc cigale}; \citealt{Boquien_2019}), with the stellar models of \citet{Bruzual_2003}, \citet{Chabrier_2003} initial mass function, \citet{Calzetti_2000} dust attenuation law and the IGM absorption of \citet{Meiksin_2006}. Simple exponential star-formation history models, ${\rm SFH}\propto exp(-t/\tau)$, were assumed, with $\tau$ values of 0.1, 0.3, 1, 2, 3, 5, 10 and 15\,Gyr and the dust attenuation, $A_V$, ranging between 0 and 2.8 in steps of 0.05. The $NUVrJ$ magnitudes and the UV luminosities were found for each source by {\sc cigale} from the best-fit SEDs. The procedure for deriving the UV slopes is described in Section\,\ref{sec:beta}, while the determination of the IRX-$\beta$ relationship functional form parameters is explained in Section\,\ref{sec:irxb}.

\subsection{UV slopes}
\label{sec:beta}

A number of methods have been used for determining the UV spectral slope $\beta$, a comprehensive review of which is presented in \citet{Rogers_2013}. Originally, \citet{Meurer_1999} calculated $\beta$ by fitting a power-law function to the ten continuum rest-frame 125-250\,nm bands of \citet{Calzetti_1994}. Most often, however, only a few photometric points are available in this wavelength range and, hence, the resulting $\beta$'s are burdened with significant uncertainties. As explained by \citet{McLure_2018}, these uncertainties can bias the stacked values of the IRX, flattening the corresponding IRX-$\beta$ relationship. Therefore, we derived the UV slopes by fitting the power-law function to the best-fit SEDs in the rest-frame wavelength range of $125-250$\,nm, where $S_\lambda\propto \lambda^\beta$. As recently shown by \citet{Morales_2024}, SED-based values exhibit lower scatter and reduced biases than those derived from photometric power-law fitting.

To estimate the UV slope uncertainties for each source, the photometry was randomly sampled from normal distributions, with the means and standard deviations corresponding to the catalog values and their associated errors and the resulting value of $\beta_i$ derived from the best-fit SED. The procedure was repeated 1000 times and the final errors were taken to be the standard deviations of the individual values of $\beta_i$.

\subsection{Stacked IRX values}
\label{sec:irx}

For about 600k star-forming galaxies in our optical/near-IR catalogs, only $\sim$2000 sources were directly detected in the S2CLS 850\,${\rm \mu m}$ survey \citep{Geach_2017}. In order to get reliable estimates of the IR luminosity, and hence the IRX, we, therefore, resorted to the well-known method of stacking (e.g., \citealt{Tomczak_2016, Leslie_2020, Merida_2023, Koprowski_2024}).  Because it is not clear how does $L_{\rm IR}$ couple with $L_{\rm UV}$, the stacked values of the IRX were taken to be equal to the median of the individual values of the infrared excess found for each source in the stack. To compute this, we read the SCUBA-2 850-${\rm \mu m}$ flux density at each source's position, converted it to the IR luminosity and divided by the corresponding UV luminosity. $L_{\rm IR}$ was found by integrating the best-fit dust emission curve of \citet{Casey_2012} between 8-1000\,${\rm \mu m}$ rest frame, with the emissivity and the mid-IR power-law slope set to 1.96 and 2.3, respectively, as per recommendations of \citet{Drew_2022}. To determine the dust temperature, we adopted the $T_{\rm d}$-$z$ relation from \citet{Koprowski_2024}, which was derived using the same sample analyzed in this study. In each bin, the corresponding stacked values of the UV slope, stellar mass and redshift were also derived by taking the median of all the individual numbers.

The impact of redshift and stellar-mass uncertainties on the stacked values of the IRX, $\beta$, $M_\ast$ and redshift was quantified using a bootstrap method. At each iteration, a mock catalog was generated by randomly drawing, with replacement, a set of sources from the original mass-complete dataset, for which a stacked value was found, following the procedure described above. The process was repeated 1000 times and the errors, $\delta_{\rm bootstrap}$, were then taken to be the median absolute deviation of the resulting simulated values. In addition, to account for uncertainties in the UV luminosity and UV slope, we performed simple Monte Carlo simulations in which, in each of 1000 runs, the values of $L_{\rm UV}$ and $\beta$ were randomly drawn from Gaussian distributions with standard deviations set to their respective $1\sigma$ uncertainties. The corresponding errors, $\delta_{\rm MC}$, were taken to be equal to the median of the individual values found in each run. The final uncertainties were then determined, where:

\begin{equation}\label{eq:dirx}
    \delta = \sqrt{\delta^2_{\rm bootstrap}+\delta^2_{\rm MC}}.
\end{equation}

\section{Analysis \& Discussion}\label{sec:disc}

\subsection{IRX-$\beta$ in bins of UV slope}\label{sec:irxb}

\begin{figure}
\centering
   \includegraphics[width=9cm]{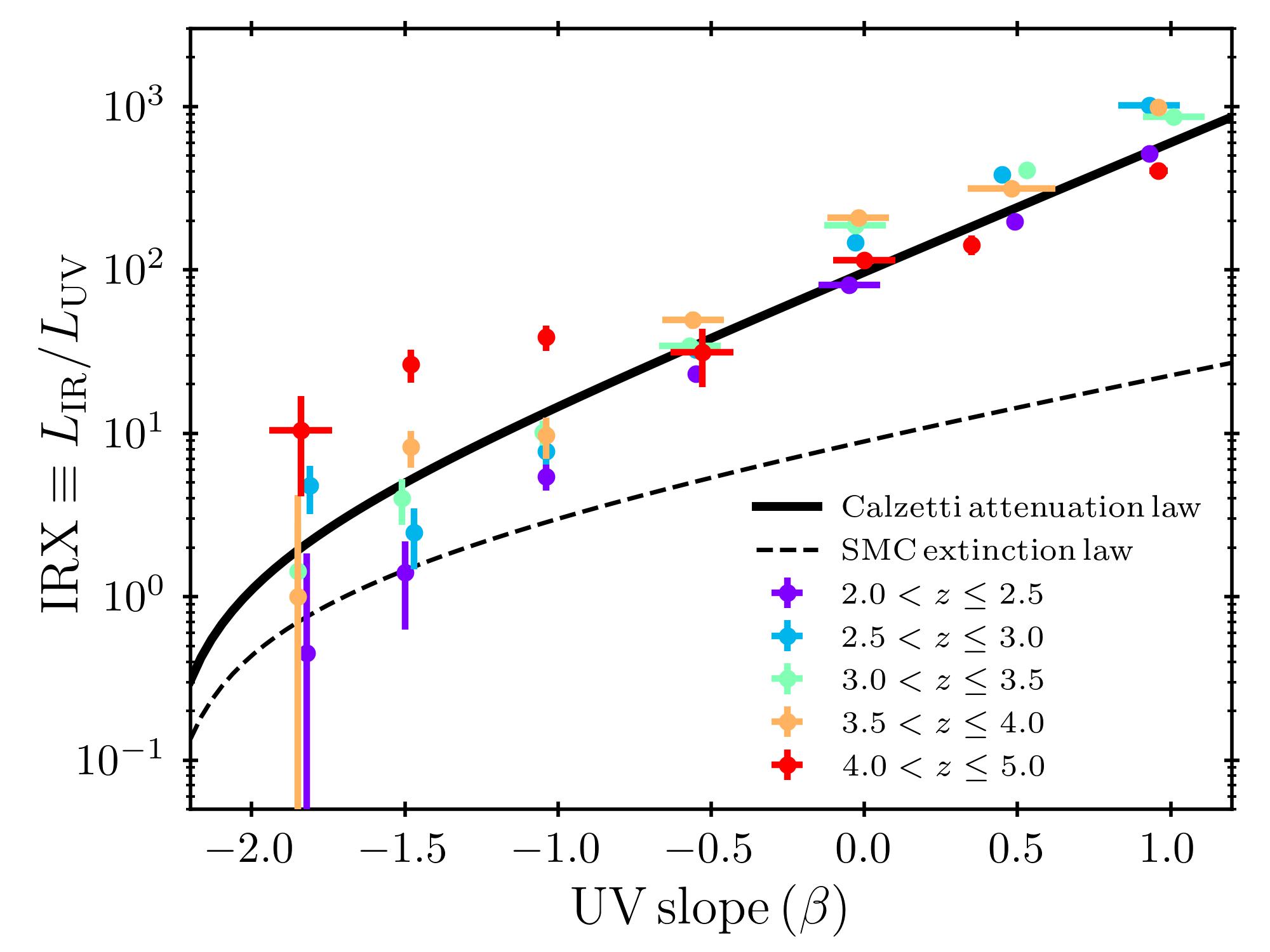}
     \caption{IRX-$\beta$ relationship for $2.0<z\leq 5.0$ sample studied in this work. The color points with error bars show the stacked values of the IRX in bins of $\beta$, summarized in Table\,\ref{tab:irxb}. The black solid line represents the best-fit functional form (Equation\,\ref{eq:sum1}), with $dA_{1600}/d\beta=1.97$ being consistent with the attenuation curve of \citet{Calzetti_2000}. The scatter at $\beta\lesssim-1$ is caused by different stellar mass completeness limits imposed at each redshift bin (see Section\,\ref{sec:irxb} for details). For reference, the SMC curve is plotted in dashed line.}
     \label{fig:irxb}
\end{figure}

In order to find the evolution of the infrared excess with the UV slope and redshift, we stacked our mass-complete sample following the procedure explained in Section\,\ref{sec:irx}. The redshift and $\beta$ bins and the mass-completeness limits, adopted from \citet{McLeod_2021}, with the corresponding IRX-$\beta$ median values are summarized in Table\,\ref{tab:irxb} and plotted in Figure\,\ref{fig:irxb} as color points with error bars.

The functional form was adopted from \citet{Meurer_1999}:

\begin{equation}\label{eq:firxb}
    {\rm IRX} = (10^{0.4A_{1600}}-1)\times B,
\end{equation}

\noindent where $A_{1600}$ is the attenuation at the rest-frame 1600\AA\, and $B$ is the ratio of two correction factors:

\begin{equation}\label{eq:b}
    B=\frac{\rm BC(1600)}{\rm BC(FIR)}.
\end{equation}

Since, in the original work of \citet{Meurer_1999}, the FIR luminosity, measured using $IRAS$ data, was defined as $L_{\rm FIR}=1.25(L_{60}+L_{100})$ and the UV luminosity was determined at the rest-frame 1600\AA, corrections were required in order to convert into bolometric luminosities. In this work, the IR luminosity is found by integrating the best-fit dust emission curve between 8-1000\,${\rm \mu m}$ (Section\,\ref{sec:irx}), so the FIR correction factor is by definition equal to 1. BC(1600), on the other hand, can be found once the intrinsic stellar emission curve is known. The attenuation at the rest-frame 1600\AA\, is defined as:

\begin{equation}\label{eq:att}
    A_{1600}=\frac{dA_{1600}}{d\beta}(\beta_{\rm obs}-\beta_{\rm int}),
\end{equation}

\noindent with $\beta_{\rm obs}$ and $\beta_{\rm int}$ being the UV slopes of the observed and the intrinsic stellar emission spectra, respectively.

The available UV-FIR photometry (Section\,\ref{sec:data}) for all the sources between redshifts 2.0 and 5.0 was median stacked in all the available FIR maps (Section\,\ref{sec:fir}) and the corresponding best-fit observed and intrinsic SEDs were found using {\sc cigale}, as explained in Section\,\ref{sec:sed}, with the slope of the dust reddening law, $dA_{1600}/d\beta$, set as a free parameter. The intrinsic stellar emission curve was then used to calculate the intrinsic UV slope and the bolometric correction factor, BC(1600):

\begin{equation}\label{eq:b0bc}
\begin{split}
    \beta_{\rm int} &= -2.30\pm 0.10, \\
    {\rm BC(1600)} &= \phantom{-}1.51\pm 0.11,
\end{split}
\end{equation}

\noindent with the resulting best-fit IRX-$\beta$ relationship functional form plotted as black solid line in Figure\,\ref{fig:irxb}. In order to estimate the errors, we performed simple Monte Carlo simulations, where in each of 1000 realizations, the photometry was randomly sampled from the Gaussian distribution with the means and standard deviations equal to the catalog's values and their errors, respectively. The best-fit SEDs were found and the bolometric correction, BC(1600), and $\beta_{\rm int}$ calculated for each run. The final errors were then set to be equal to the median of all the individual values. The best-fit slope, $dA_{1600}/d\beta=1.97$, is consistent with the \citet{Calzetti_2000} reddening law, where the values of:

\begin{equation}\label{eq:dadbl}
    dA_{1600}/d\beta=1.97\,\,\,\,{\rm and} \,\,\,\,dA_{1600}/d\beta=0.91,
\end{equation}

\noindent for the Calzetti attenuation law and SMC extinction curve were assumed, respectively \citep{McLure_2018}. Accordingly, our best-fit IRX–$\beta$ relation is hereafter referred to as a Calzetti-like attenuation law.

\begin{figure}
\centering
   \includegraphics[width=9cm]{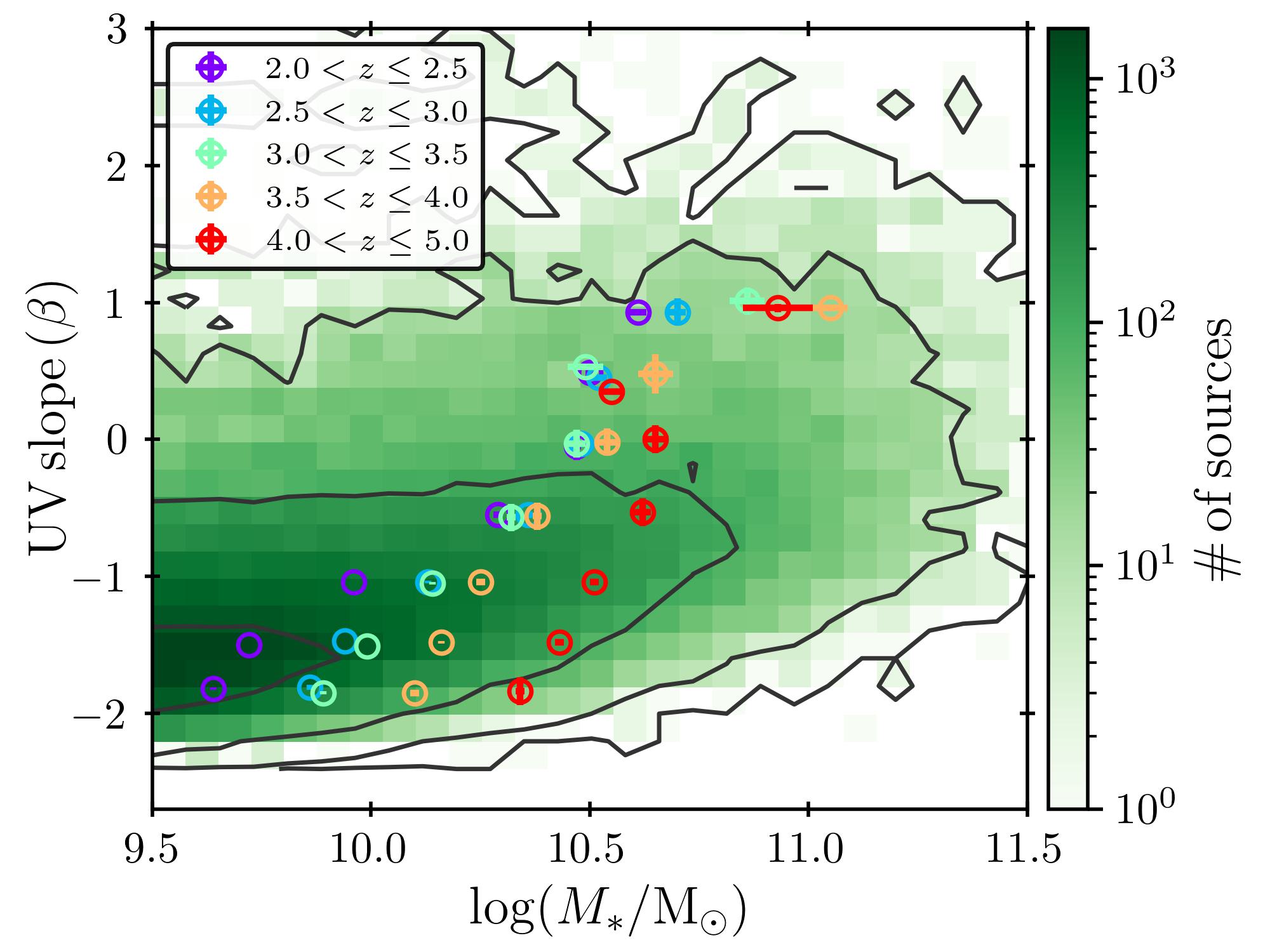}
     \caption{Relationship between the UV slope and the stellar mass for the sample presented in Figure\,\ref{fig:irxb}. Green background image shows a 2D histogram of the $\beta$-$M_\ast$ distribution, with 1, 10, 100 and 1000 sources contours displayed with black curves. It can be seen that due to different mass-completeness limits imposed at each redshift bin (Table\,\ref{tab:irxb}), the median stellar masses at bluest $\beta$ bins increase with redshift, which in turn drives the scatter in Figure\,\ref{fig:irxb} (see Section\,\ref{sec:irxb} for details).}
     \label{fig:betam0}
\end{figure}

It can be seen in Figure\,\ref{fig:irxb} that at $\beta>-1.0$ our IRX-$\beta$ relationship is in agreement with the \citet{Calzetti_2000} attenuation curve across all redshift bins. At bluer $\beta$'s, however, the values of the IRX seem to be increasing with redshift. In Figure\,\ref{fig:betam0} the median values of the UV slope and stellar mass, corresponding to the stacked data of Figure\,\ref{fig:irxb}, color-coded with redshift are plotted. The green background image shows a 2D histogram of the $\beta_{\rm obs}$–$M_\ast$ distribution, with the black contours corresponding to the values of 1, 10, 100 and 1000. It can be seen that at bluer $\beta$'s the median stellar masses are significantly larger at higher redshifts, reflecting different stellar mass completeness limits imposed in each redshift bin (Table\,\ref{tab:irxb}). This aligns with recent results indicating that IRX correlates with stellar mass, such that, at fixed $\beta$, systems with higher stellar masses exhibit larger values of the infrared excess (e.g., \citealt{Alvarez_2019, Bouwens_2020}). In order to investigate this further, in the next section we stack individual values of the infrared excess in bins of the stellar mass.

\subsection{IRX-$\beta$ in bins of stellar mass}\label{sec:mirx}

\begin{figure}
\centering
   \includegraphics[width=9cm]{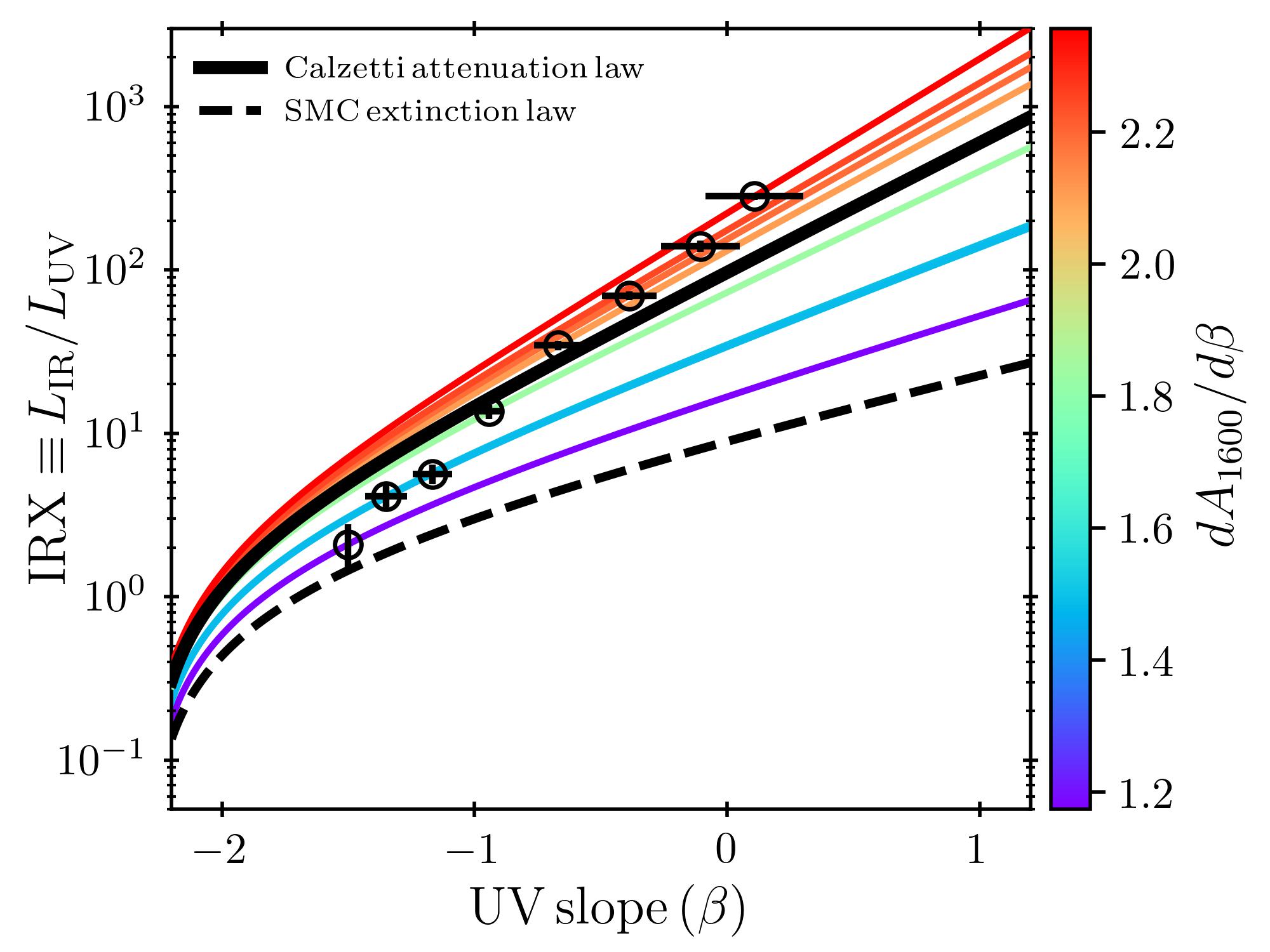}
     \caption{IRX-$\beta$ relationship in bins of stellar mass. The black circles show the stacked data (Table\,\ref{tab:irxbm}), with a clear correlation between the stellar mass and the slope of the reddening law, $dA_{1600}/d\beta$. Adopting Equation\,\ref{eq:sum1}, with $dA_{1600}/d\beta$ set as free parameter, best-fit values for the reddening slope were found (last column in Table\,\ref{tab:irxbm}), with the corresponding correlation with the stellar mass derived in Section\,\ref{sec:mirx} and plotted in Figure\,\ref{fig:dadbm}. For reference, this work’s best-fit relationship (consistent with the Calzetti curve), together with the SMC curve are plotted in solid and dashed lines, respectively.}
     \label{fig:irxbm}
\end{figure}

\begin{table*}
\begin{footnotesize}
\begin{center}
\caption{Physical properties for the $2.0<z\leq 5.0$ sample stacked in Section\,\ref{sec:mirx} and plotted in Figure\,\ref{fig:irxbm}.}\label{tab:irxbm}
\setlength{\tabcolsep}{4 mm}
\begin{tabular}{ccccc}
\hline\hline
& $\beta$ & $M_\ast$ & log(IRX) & $dA_{1600}/d\beta$ \\ 
\hline
$\phantom{1}9.50< {\rm log}(M_\ast/{\rm M_\odot}) \leq \phantom{1}9.75$ & $-1.50 \pm 0.00$ & $\phantom{1}9.62$ & $0.32^{+0.09}_{-0.12}$ & $1.17 \pm 0.13$ \\ 
$\phantom{1}9.75< {\rm log}(M_\ast/{\rm M_\odot}) \leq 10.00$ & $-1.35 \pm 0.01$ & $\phantom{1}9.86$ & $0.61^{+0.05}_{-0.05}$ & $1.50 \pm 0.07$ \\ 
$10.00< {\rm log}(M_\ast/{\rm M_\odot}) \leq 10.25$ & $-1.17 \pm 0.01$ & $10.11$ & $0.75^{+0.04}_{-0.05}$ & $1.49 \pm 0.06$ \\ 
$10.25< {\rm log}(M_\ast/{\rm M_\odot}) \leq 10.50$ & $-0.94 \pm 0.00$ & $10.36$ & $1.13^{+0.04}_{-0.04}$ & $1.84 \pm 0.05$ \\ 
$10.50< {\rm log}(M_\ast/{\rm M_\odot}) \leq 10.75$ & $-0.67 \pm 0.02$ & $10.61$ & $1.54^{+0.03}_{-0.03}$ & $2.11 \pm 0.04$ \\ 
$10.75< {\rm log}(M_\ast/{\rm M_\odot}) \leq 11.00$ & $-0.39 \pm 0.02$ & $10.85$ & $1.84^{+0.03}_{-0.03}$ & $2.18 \pm 0.03$ \\ 
$11.00< {\rm log}(M_\ast/{\rm M_\odot}) \leq 11.25$ & $-0.11 \pm 0.03$ & $11.09$ & $2.15^{+0.04}_{-0.04}$ & $2.25 \pm 0.04$ \\ 
$11.25< {\rm log}(M_\ast/{\rm M_\odot}) \leq 11.50$ & $\phantom{-}0.11 \pm 0.04$ & $11.32$ & $2.45^{+0.03}_{-0.03}$ & $2.36 \pm 0.04$ \\ 
\hline
\end{tabular}
\end{center}
\end{footnotesize}
\end{table*}

Following the procedure explained in Section\,\ref{sec:irx}, the stacked median values of the infrared excess were found in bins of the stellar mass for all the sources between redshifts 2.0 and 5.0, where in each stellar mass bin the corresponding UV slope was assumed to be the median of all the individual $\beta$'s. The results of this exercise are presented in Figure\,\ref{fig:irxbm} with black open circles and summarized in Table\,\ref{tab:irxbm}. For reference, the best-fit functional form of the IRX-$\beta$ relation found in Section\,\ref{sec:irxb} (consistent with the \citealt{Calzetti_2000} curve) and the SMC-like relation are plotted as black solid and dashed lines, respectively. It can be clearly seen that the position of a given IRX-$\beta$ stacked data point relative to the best-fit relation correlates with the median stellar mass. 

As explained in a number of recent works (e.g., \citealt{Narayanan_2018, Koprowski_2018, Salim_2020}), the scatter in the IRX-$\beta$ plane is mainly caused by the variations in the slopes of the underlying attenuation laws, where galaxies affected by flatter (grayer) curves are shifted towards bluer UV slopes. \citet{Salim_2020} show that the attenuation curves slopes correlate strongly with effective optical opacity (i.e. the dust column density), with dusty and more massive galaxies having, on average, grayer curves. They attribute this relationship primarily to the combined effects of radiative transfer processes (scattering and absorption) and local geometry (clumpiness of dust). In this framework, a fraction of the rest-frame UV light, originating from young massive stars, escapes the galaxy through gaps in the dust distribution and via scattering processes, causing the effective attenuation curve to flatten. 

It has been shown that galaxies at high redshifts exhibit a large variety of dust attenuation laws (e.g., \citealt{Kriek_2013, LoFaro_2017, Salmon_2016, Cullen_2018}). \citet{Alvarez_2019} investigated a sample of high-redshift LBGs and found more massive sources to be affected by grayer attenuation curves. Similar trend was found by \citet{Bouwens_2020}. More recently, \citet{Shivaei_2025} investigated a spectroscopic sample from three JWST/NIRCam grism surveys and found the slopes of the attenuation curves to become shallower at higher values of the dust attenuation. Given that the dust attenuation is positively correlated with the stellar mass (i.e. more massive galaxies tend to be more dusty; Figure\,\ref{fig:betam0}, but also \citealt{McLure_2018, Bouwens_2020}), higher-mass sources are systematically expected to be characterized by grayer attenuation curves.

\begin{figure}
\centering
   \includegraphics[width=9cm]{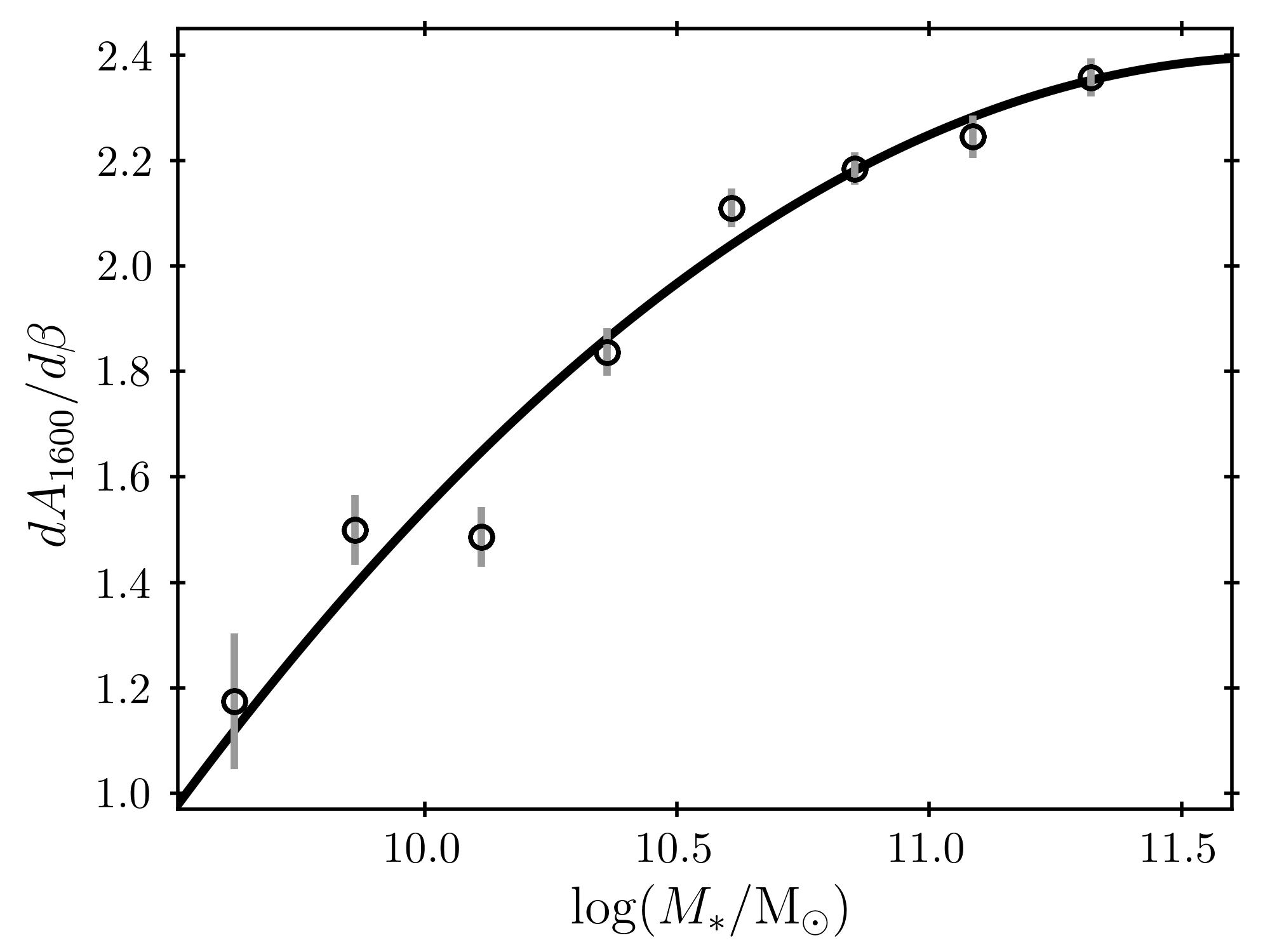}
     \caption{Relationship between the dust reddening slope, $dA_{1600}/d\beta$, and the stellar mass for the sample plotted in Figure\,\ref{fig:irxbm}, where a clear correlation can be seen, with more massive systems being characterized by flatter attenuation curves. The functional form of the relation is derived and discussed in Section\,\ref{sec:mirx} and summarized in Equation\,\ref{eq:sum1}.}
     \label{fig:dadbm}
\end{figure}

We, therefore, proceed to establish the functional form of the relationship between the stellar mass and the slope of the underlying attenuation law using data from Figure\,\ref{fig:irxbm}. For practical reasons, the attenuation curve for a given stellar mass bin is quantified with the dust reddening slope of Equation\,\ref{eq:att}, $dA_{1600}/d\beta$. A steeper attenuation curve leads to greater relative attenuation toward the edges of the UV wavelength range over which the UV slope $\beta$ is measured (i.e., 125–250 nm in the rest frame). That means that, for a fixed attenuation, the steeper curve will produce redder $\beta$, shifting the galaxy to the right on the IRX-$\beta$ plane. Consequently, steeper attenuation curves cause $\beta$ to redden more rapidly as attenuation increases, giving rise to shallower reddening slopes, $dA_{1600}/d\beta$.

Assuming the median intrinsic stellar SED parameters, BC(1600) and $\beta_{\rm int}$, found in Section\,\ref{sec:irxb} for our $2.0<z\leq 5.0$ sample, $dA_{1600}/d\beta$ was established for each IRX-$\beta$ stacked point in Figure\,\ref{fig:irxbm} from Equation\,\ref{eq:firxb} using the $\chi^2$ minimization method. The resulting values are summarized in Table\,\ref{tab:irxbm} and plotted in Figure\,\ref{fig:dadbm} with open circles, with the corresponding functional forms of the IRX-$\beta$ relation depicted in Figure\,\ref{fig:irxbm} with solid color lines. The correlation between the slope, $dA_{1600}/d\beta$, and the stellar mass was then found using the $\chi^2$ minimization method (solid curve in Figure\,\ref{fig:dadbm}), where:

\begin{equation}\label{eq:dadbm}
\begin{split}
    dA_{1600}/d\beta & = a\mathcal{M}^2+b\mathcal{M}+c, \\
    a & = -0.30\pm 0.07, \\
    b & = \phantom{-}7.06\pm 1.40, \\
    c & = -38.8\pm 7.5,
\end{split}
\end{equation}

\noindent with $\mathcal{M}={\rm log}(M_\ast/{\rm M_\odot})$ and the errors estimated from the corresponding co-variance matrix. Therefore, the infrared excess becomes a function of both the UV slope and the stellar mass, where the later enters the equation through its effect on the slope of the reddening law, $dA_{1600}/d\beta$, with the corresponding functional forms summarized in Table\,\ref{tab:firxbm}. The comparison between the functional form found in this work and the ones determined in \citet{Alvarez_2019} and \citet{Bouwens_2020} are presented in Sections\,\ref{sec:irxm} and \ref{sec:irxmcomp}.

\subsection{IRX-$\beta$ -- comparison to previous studies}\label{sec:irxbcomp}

\begin{figure}
\centering
   \includegraphics[width=9cm]{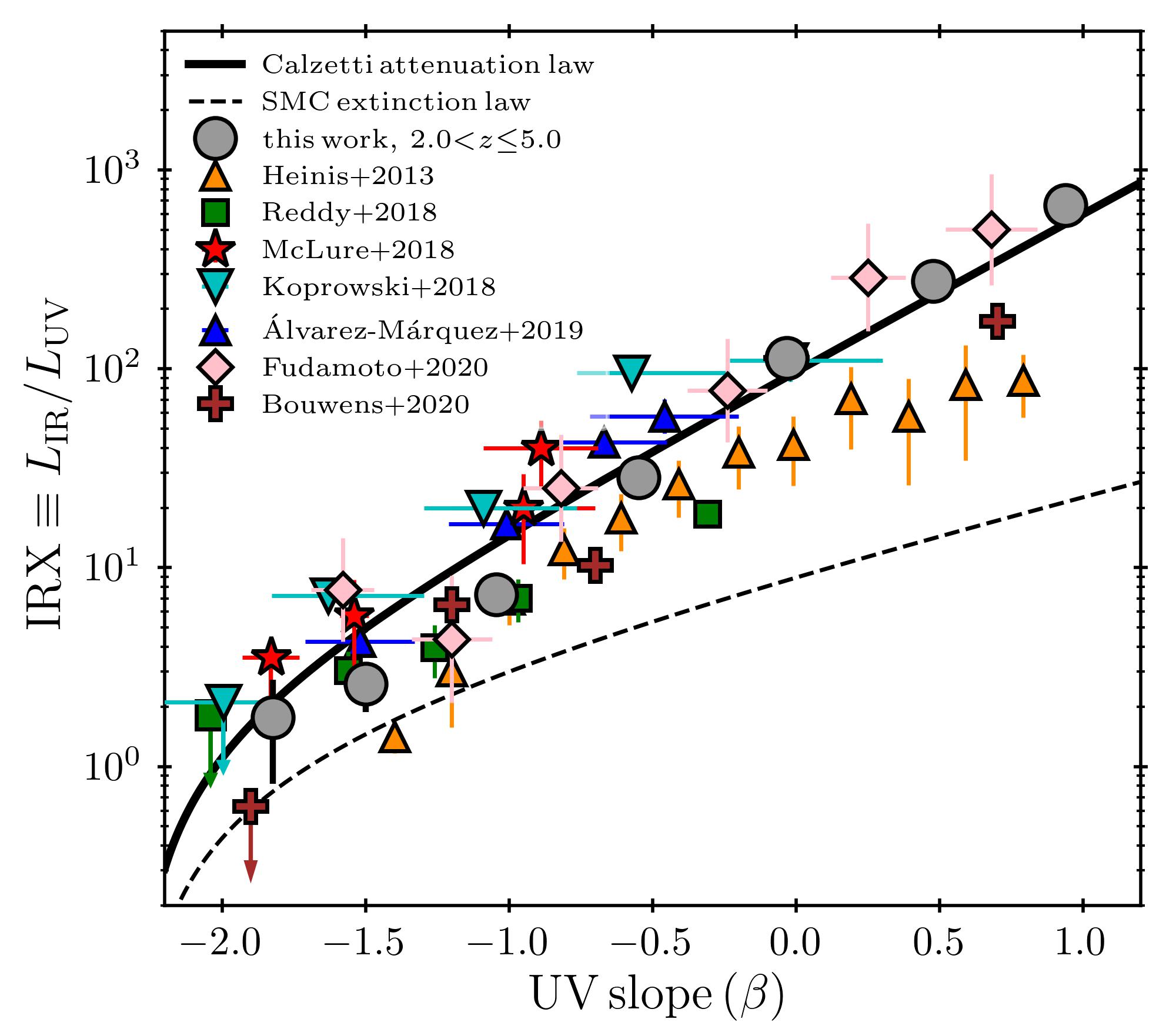}
     \caption{IRX-$\beta$ relationship derived for the $2.0<z \leq 5.0$ sample of this work (gray circles; Table\,\ref{tab:irxb1}). Our median data lying slightly below the best-fit curve at blue UV slopes is driven by variations in the corresponding median stellar masses. We compare our data with the recent literature results of \citet{Heinis_2013, Reddy_2018, McLure_2018, Koprowski_2018, Alvarez_2019, Fudamoto_2020} and \citet{Bouwens_2020}, with apparent inconsistencies discussed in Section\,\ref{sec:irxbcomp}. For reference, this work's best-fit relationship and the SMC-like curve are plotted in solid and dashed lines, respectively.}
     \label{fig:irxbcomp}
\end{figure}

We plot our stacked data in Figure\,\ref{fig:irxbcomp}. The gray circles represent median values of the infrared excess resulting from stacking all the sources between redshifts 2.0 and 5.0 (Table\,\ref{tab:irxb1}), adopting stellar mass limits from Table\,\ref{tab:irxb}. For reference, this work's best-fit relationship (consistent with the Calzetti curve), together with the SMC curve are plotted in solid and dashed lines, respectively. It can be seen that our median data is still slightly below the best-fit curve at blue UV slopes. This is caused by the variations in the corresponding stellar masses, with bluer bins having slightly lower median values of $M_\ast$.

For comparison, we also plot in Figure\,\ref{fig:irxbcomp} the recent literature results of \citet{Heinis_2013, Reddy_2018, McLure_2018, Koprowski_2018, Alvarez_2019, Fudamoto_2020} and \citet{Bouwens_2020}. The stacked data found in this work seem to be in a reasonable agreement with the works of \citet{McLure_2018, Koprowski_2018, Alvarez_2019} and \citet{Fudamoto_2020}, with the slightly elevated values found in \citet{Koprowski_2018} most likely caused by the higher dust temperature assumed when deriving their IR luminosities. The numbers presented in \citet{Heinis_2013, Reddy_2018} and \citet{Bouwens_2020} are slightly below our stacked data. However, due to the variety of methods employed to measure both UV slopes and IRX, the direct comparison between these works is difficult. 

As mentioned in Section\,\ref{sec:beta}, different ways of determining the UV slope can cause the resulting IRX-$\beta$ relationship to flatten \citep{McLure_2018}. Once $\beta$'s are found, one needs to ensure that the stacked sample is complete. Given the large beam sizes of the {\it Herschel} SPIRE data, the clustering effects must also be considered, where one needs to ensure that multiple sources falling into one SPIRE beam are not counted multiple times, causing the stacked fluxes to be overestimated. In this work we followed \citet{Koprowski_2024}, where median SPIRE fluxes were modeled with the two-component light profile, comprising the actual beam and the contribution from background sources. In addition, we made use of the 850\,${\rm \mu m}$ SCUBA-2 data, which not only has higher resolution than SPIRE but also allowed us to constrain the Rayleigh–Jeans tail of the dust emission curve. \citet{Reddy_2018} estimated the so-called bias factor, defined as the ratio of the simulated to recovered fluxes, using simulated maps, while the calculations presented in \citet{Heinis_2013} were based on the assumption that the overestimation of the stacked FIR fluxes is proportional to the sample's angular correlation function. The different assumptions behind the shape of the dust emission curve (the shape, dust temperature, emissivity, etc.) will have an additional impact on the inferred IR luminosities. Furthermore, since it is not clear how the $L_{\rm UV}$ couples with the $L_{\rm IR}$, stacking individual values of the IR luminosities and dividing by the corresponding $L_{\rm UV}$ produces slightly different numbers than stacking individual values of the infrared excess. Finally, as explained in Section\,\ref{sec:mirx}, the infrared excess at any given UV slope will also vary with the stellar mass.

\subsection{IRX-$M_\ast$}\label{sec:irxm}

\begin{figure}
\centering
   \includegraphics[width=9cm]{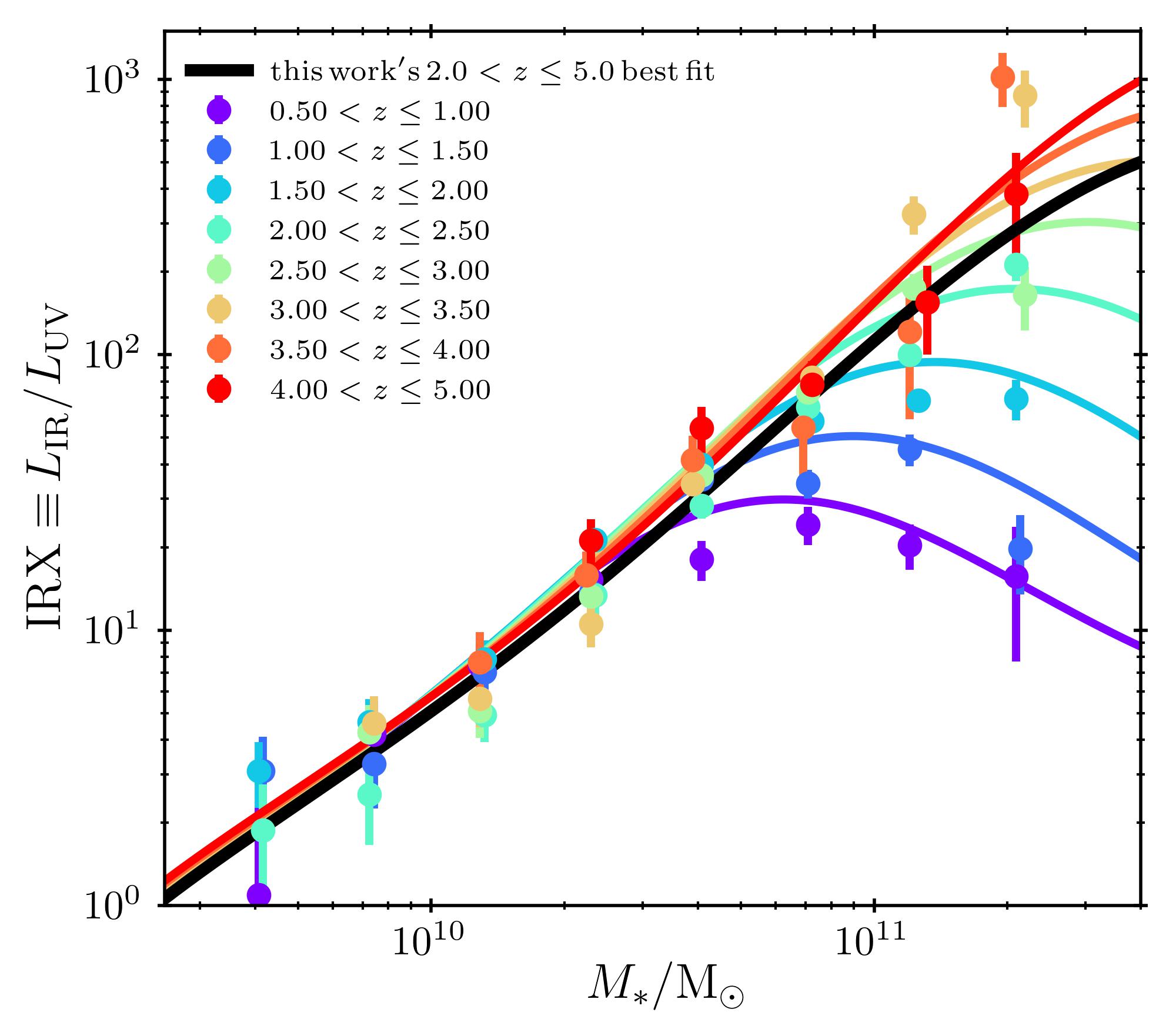}
     \caption{IRX values median-stacked in bins of $M_\ast$ and redshift (color points with error bars; Table\,\ref{tab:irxm}). The black line shows the best-fit functional form for the $2.0<z\leq 5.0$ sample of Figure\,\ref{fig:irxbm}, while the color solid lines depict the functional forms found at the individual redshift bins, as explained in detail in Section\,\ref{sec:irxm}.}
     \label{fig:irxm1}
\end{figure}

\begin{figure}
\centering
   \includegraphics[width=9cm]{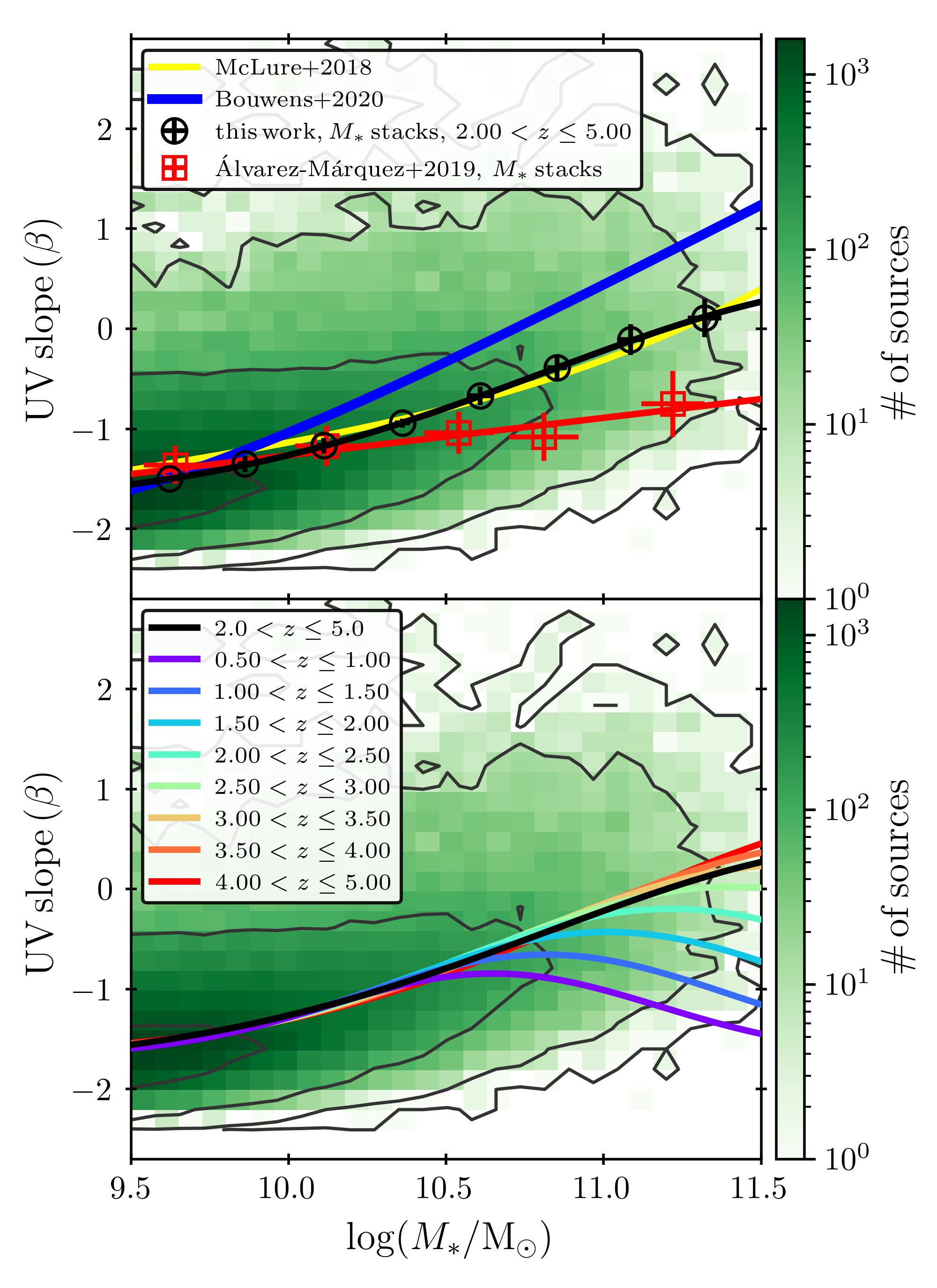}
     \caption{The relationship between the observed UV slope and stellar mass for the $2.0<z\leq 5.0$ sample of this work. Green background image represents a 2D histogram of the $\beta$-$M_\ast$ distribution, with 1, 10, 100 and 1000 sources contours displayed with thin black curves. \textbf{Top panel.} Black circles with error bars show median values of $\beta$ in bins of $M_\ast$ (Table\,\ref{tab:irxbm}), with the thick black solid curve depicting the best-fit functional form of Equation\,\ref{eq:fbm}. For comparison, the relationships found in \citet{McLure_2018, Alvarez_2019} and \citet{Bouwens_2020} are shown. \textbf{Bottom panel.} Black curve as in the top panel with the color curves determined for individual redshift bins, following the procedure explained in Section\,\ref{sec:irxm}.}
     \label{fig:betam}
\end{figure}

\begin{table*}
\begin{footnotesize}
\begin{center}
\caption{Functional form of the IRX-$\beta$-$M_\ast$-$z$ relationship.}
\label{tab:firxbm}
\setlength{\tabcolsep}{2 mm} 
\begin{tabular}{c}
\hline
\hline
IRX-$\beta$ relationship found in Section\,\ref{sec:irxb} \\
\hline
\begin{minipage}{0.33\linewidth}
\centering
\begin{equation}
\begin{aligned}
{\rm IRX} &= (10^{0.4A_{1600}}-1)\times B \\
A_{1600} &= \frac{dA_{1600}}{d\beta} \, (\beta_{\rm obs}-\beta_{\rm int}) \\
B &= 1.51\pm 0.11 \\
\beta_{\rm int} &= -2.30\pm 0.10
\end{aligned}
\label{eq:sum1}
\end{equation}
\end{minipage} \\
\hline
$dA_{1600}/d\beta$ dependence on the stellar mass found in Section\,\ref{sec:mirx} \\
\hline
\begin{minipage}{0.33\linewidth}
\centering
\begin{equation}
\begin{aligned}
\frac{dA_{1600}}{d\beta} &= a\mathcal{M}^2+b\mathcal{M}+c \\
\mathcal{M} &= {\rm log}(M_\ast/{\rm M_\odot}) \\
a & = -0.30\pm 0.07, \\
b & = 7.06\pm 1.40, \\
c & = -38.8\pm 7.5
\end{aligned}
\label{eq:sum2}
\end{equation}
\end{minipage} \\
\hline
$\beta_{\rm obs}$-$M_\ast$ relationship evolution with redshift established in Section\,\ref{sec:irxm} \\
\hline
\begin{minipage}{0.33\linewidth}
\centering
\begin{equation}
\begin{aligned}
\beta_{\rm obs} &= a+b\times {\rm exp}\left(-\frac{(\mathcal{M}-c)^2}{2d^2}\right) \\
\mathcal{M} &= {\rm log}(M_\ast/{\rm M_\odot}) \\
a &= -1.78\pm 0.03 \\
b & = (0.45\pm 0.04)\times z+ (0.56\pm 0.07) \\
c & = (0.40\pm 0.04)\times z+ (10.33\pm 0.06) \\
d & = (0.16\pm 0.03)\times z+ (0.44\pm 0.07)
\end{aligned}
\label{eq:sum3}
\end{equation}
\end{minipage} \\
\hline
\end{tabular}
\end{center}
\end{footnotesize}
\end{table*}

Since measurements of the UV slope, $\beta$, are often burdened with significant uncertainties, causing additional scatter in the corresponding IRX values, it is often preferable to express the infrared excess as a function of stellar mass. We, therefore, stacked our full $0.5<z\leq 5.0$ sample in bins of redshift and $M_\ast$, as summarized in Table\,\ref{tab:irxm}, with the resulting median values plotted in Figure\,\ref{fig:irxm1} with color points. In order to find the functional form, we started with the IRX-$\beta$ relationship found in Section\,\ref{sec:irxb} (Equations\,\ref{eq:sum1} summarized in Table\,\ref{tab:firxbm}), with the functional form of the stellar mass evolution of the reddening law, $dA_{1600}/d\beta$, found in Section\,\ref{sec:mirx} (Equations\,\ref{eq:sum2} summarized in Table\,\ref{tab:firxbm}). To express the IRX as a sole function of stellar mass, we converted the observed UV slope, $\beta_{\rm obs}$, into stellar mass, $M_\ast$. This was done using data corresponding to the stacked points of Figure\,\ref{fig:irxbm}, plotted as black open circles in the top panel of Figure\,\ref{fig:betam}, by fitting a Gaussian function (black solid line in Figure\,\ref{fig:betam}), using non-linear least squares with the errors estimated from the corresponding co-variance matrix. The resulting relation can be written as:

\begin{equation}\label{eq:fbm}
    \beta=a+b\times {\rm exp}\left(-\frac{(\mathcal{M}-c)^2}{2d^2}\right),
\end{equation}

\noindent with $\mathcal{M}={\rm log}(M_\ast/{\rm M_\odot})$ and:

\begin{equation}\label{eq:fbmpars}
\begin{split}
    a & = \phantom{}-1.78\pm 0.03 \\
    b & = \phantom{-1}2.24\pm 0.14 \\
    c & = \phantom{-}12.00\pm 0.13 \\
    d & = \phantom{-1}1.16\pm 0.07. \\
\end{split}
\end{equation}

For comparison, in the top panel of Figure\,\ref{fig:betam}, the similar relationships found in \citet{McLure_2018, Alvarez_2019} and \citet{Bouwens_2020} are plotted with yellow, red and blue solid lines, respectively. An excellent agreement between our relationship and the one found in \citet{McLure_2018} can clearly be seen. The results of \citet{Alvarez_2019} are based on a sample of $2.5<z<3.5$ COSMOS field Lyman Break Galaxies. Since LBGs are selected at the rest-frame UV bands, a fraction of red sources will likely be missed, underestimating the resulting stacked values of the UV slopes. Data investigated in \citet{Bouwens_2020}, on the other hand, shows elevated $\beta$'s at high stellar masses. The source of this inconsistency can be attributed to a slightly different method of deriving the best-fit $\beta$-$M_\ast$ relation. In this work, the stacked points in the top panel of Figure\,\ref{fig:betam} were found by taking median values of individual UV slopes in each stellar mass bin, with the best-fit function found from non-linear least squares (similarly to \citealt{McLure_2018}). In \citet{Bouwens_2020}, however, the number of sources on each side of the best-fit function was minimized, which produces slightly different results. 

By combining Equation\,\ref{eq:fbm} with Equations\,\ref{eq:sum1} and \ref{eq:sum2}, we obtain the best-fit IRX–$M_\ast$ relationship shown as the black solid line in Figure\,\ref{fig:irxm1}. While it is in agreement with the low-mass stacked data, at higher stellar masses large inconsistencies can clearly be seen. Moreover, the disagreement between the best-fit function and the stacked data seems to increase towards lower redshifts, where the IRX-$M_\ast$ data exhibits a clear turnover. This behavior, which has been detected in previous works (e.g., \citealt{Whitaker_2014, Pannella_2015}), can also be seen in the shape of the star-forming galaxies main sequence (e.g., \citealt{Tomczak_2016, Popesso_2023, Koprowski_2024}). As discussed by \citet{Daddi_2022}, the turnover in the star-forming main sequence can be caused by phasing out of the cold gas accretion. At this turnover mass, supersonic shocks heat infalling molecular gas and halt cold accretion, effectively starving the galaxy of the fuel needed for sustained, dust-obscured star formation. Since, as explained in \citet{Daddi_2022}, the cold gas accretion is more efficient at higher redshifts, through cold collimated streams traveling along the dark matter filaments, the turnover of the main sequence, and hence the infrared excess, shifts towards higher masses at high redshifts, consistent with the Figure\,\ref{fig:irxm1} data. This physical transition should result in a `blueing' of the UV slope and a corresponding decrease in IRX, as the lack of new gas prevents the further accumulation of dust in these massive systems. 

We, therefore, expect the $\beta$-$M_\ast$ relation of Figure\,\ref{fig:betam} to also exhibit a turnover at high stellar masses, mimicking the behavior of the IRX-$M_\ast$ stacked data plotted in Figure\,\ref{fig:irxm1}. Since UV slopes cannot be reliably established using our data at redshifts $\lesssim2$, where the IRX-$M_\ast$ turnover is most prominent, we derived the corresponding redshift evolution of the $\beta$-$M_\ast$ relation indirectly. IRX-$M_\ast$ relation of this work is derived from the best-fit IRX-$\beta$ curve (Eq.\,\ref{eq:sum1}), with the stellar mass evolution of the reddening law given by Eq.\,\ref{eq:sum2}, and $\beta$ translated into $M_\ast$ using the appropriate form of the $\beta$-$M_\ast$ relationship. Motivated by the arguments presented above, we assume that the position and redshift evolution of the high-mass IRX-$M_\ast$ turnover is driven by the evolution of the corresponding $\beta$-$M_\ast$ curve. We, therefore, fitted the stacked IRX-$M_\ast$ data (color points in Figure\,\ref{fig:irxm1}) with Equations\,\ref{eq:sum1} and \ref{eq:sum2}, allowing the $\beta$-$M_\ast$ relation of Eq.\,\ref{eq:fbm} to vary. The derived best-fit evolution for the free parameters of the $\beta$-$M_\ast$ function, found using non-linear least squares, is given by:

\begin{equation}\label{eq:fbmparsz}
\begin{split}
    a & = \phantom{}-1.78\pm 0.03 \\
    b & = (0.45\pm 0.04)\times z+ (0.56\pm 0.07) \\
    c & = (0.40\pm 0.04)\times z+ (10.33\pm 0.06) \\
    d & = (0.16\pm 0.03)\times z+ (0.44\pm 0.07), \\
\end{split}
\end{equation}

\noindent which we summarized in Equation\,\ref{eq:sum3} of Table\,\ref{tab:firxbm} and plotted in the bottom panel of Figure\,\ref{fig:betam} with color solid lines. The corresponding evolution of the best-fit IRX-$M_\ast$ relationship for our stacked median data are depicted with solid color lines in Figure\,\ref{fig:irxm1}.

\subsection{IRX-$M_\ast$ -- comparison to previous studies}\label{sec:irxmcomp}

\begin{figure*}
\centering
   \includegraphics[width=18cm]{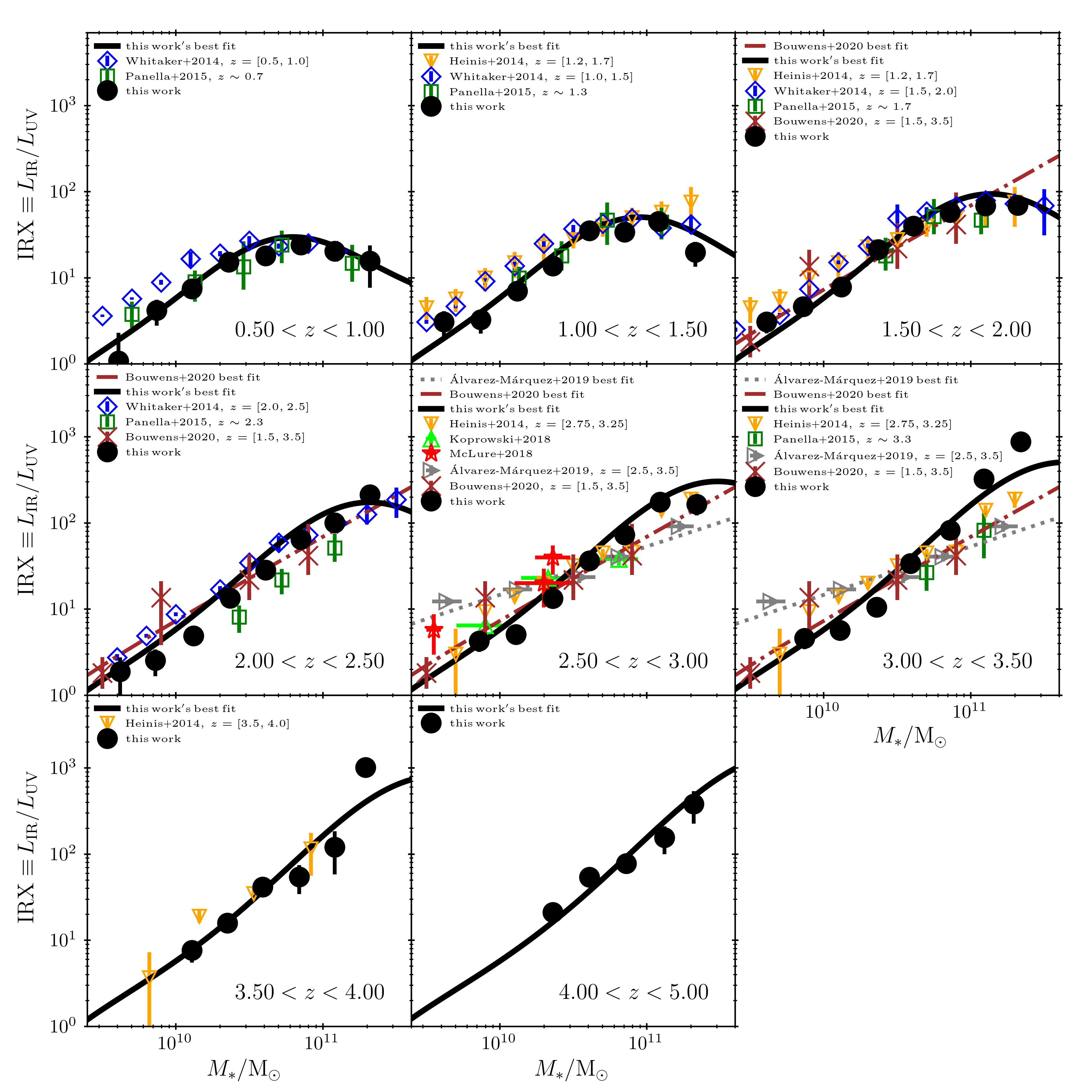}
     \caption{IRX-$M_\ast$ relationship for the $0.5<z\leq 5.0$ sample of this work. The black circles with error bars show IRX values median-stacked in bins of $M_\ast$ (Table\,\ref{tab:irxm}; same as color points in Figure\,\ref{fig:irxm1}). The black curves show corresponding best-fit functional forms summarized in Table\,\ref{tab:firxbm} (same as color curves in Figure\,\ref{fig:irxm1}). For comparison, recent literature data of \citet{Whitaker_2014, Heinis_2014, Pannella_2015, McLure_2018, Koprowski_2018, Alvarez_2019} and \citet{Bouwens_2020} are also plotted, the discussion of which is presented in Section\,\ref{sec:irxmcomp}.}
     \label{fig:irxm}
\end{figure*}

The comparison of our data with the recent literature results of \citet{Whitaker_2014, Heinis_2014, Pannella_2015, McLure_2018, Koprowski_2018, Alvarez_2019} and \citet{Bouwens_2020} is shown in Figure\,\ref{fig:irxm}. Our $0.5<z\leq 5.0$ median-stacked values are presented with black filled circles, with the best-fit functional forms, produced by combining Equations\,\ref{eq:sum1}, \ref{eq:sum2} and \ref{eq:sum3} (summarized in Table\,\ref{tab:firxbm}) depicted with black solid lines. A good agreement with most of the works can be seen, with the exception of the low-redshift \citet{Whitaker_2014} and \citet{Heinis_2014} data and the results of \citet{Alvarez_2019}. A clear high-mass turnover at low redshifts, discussed in previous section, was also detected in the works of \citet{Whitaker_2014, Heinis_2014} and \citet{Pannella_2015}, with \citet{Whitaker_2014} and \citet{Heinis_2014} finding somewhat elevated values of the IRX at low stellar masses. As explained in Section\,\ref{sec:irxbcomp}, these may arise from a number of different assumptions, most likely our choice of lower dust temperature at low redshifts, guided by the $T_{\rm d}$-$z$ relation of \citealt{Koprowski_2024}. 

The IRX-$M_\ast$ data found in \citet{Alvarez_2019} follows a significantly flatter relationship. As the authors explain in \citet{Alvarez_2019}, the inconsistencies at low stellar masses are most likely caused by the incompleteness of their LBGs sample in terms of $M_\ast$. Assuming the dependence of the stellar mass on the IRAC 3.6\,${\rm \mu m}$ flux density, the mass completeness at ${\rm log}(M_\ast/{\rm M_\odot})<10$ is between 80\% and 50\%, with lower-IRX sources missed from the sample. The IRX-$M_\ast$ data and the corresponding functional form found in \citet{Bouwens_2020} is in a very good agreement with our results, with some inconsistencies at high stellar masses. Between redshifts 3.0 and 4.0 at the highest mass bins our analysis produced somewhat elevated values of the IRX. We attribute these discrepancies to the fact that we were not able to remove all the starbursts from our sample. As explained in Section\,\ref{sec:sample}, starbursts were identified using the available ALMA data, which is not only incomplete in the COSMOS field, but is also biased against detecting sources with low dust masses and high dust temperatures. Since starbursts are known to exhibit very high dust content (e.g., \citealt{Koprowski_2018}), their presence in the high-mass sample is expected to boost the resulting IRX values.

\section{Summary}
\label{sec:sum}

In this work we investigated a $K$-band selected sample of $\sim$100000 star-forming galaxies from the UDS and COSMOS fields, spanning a combined $\sim$2 deg$^2$ with extensive UV–near-IR coverage and FIR/sub-mm imaging from {\it Herschel} and JCMT. We established the dependence of IRX on UV slope, stellar mass and redshift out to $z$$\sim$5 and to provide functional prescriptions suitable for dust correction of high-redshift galaxy samples. We adopted the available photometric redshifts and stellar masses from \citet{McLeod_2021} and removed quiescent systems and ALMA-identified starbursts where possible. IR luminosities were obtained via stacking at 850\,${\rm \mu m}$ adopting \citet{Casey_2012} SEDs and a redshift-dependent dust temperature relation from \citet{Koprowski_2024}. UV slopes were derived from SED fits in the rest-frame 125–250 nm range and the IRX computed as the median of individual IRX values. The main results can be summarized as follows:

\begin{enumerate}[label=(\roman*),wide]

\item{Stacking in the UV slope bins reveals that, at $\beta \gtrsim -1$, the IRX-$\beta$ relation is consistent with a Calzetti-like attenuation law, with a best-fit dust-reddening slope $dA_{1600}/d\beta = 1.97$ and the intrinsic UV slope $\beta_{\rm int} = -2.30$. At bluer UV slopes, however, the stacked values of the IRX increase with redshift. We found this systematic to be a direct consequence of different mass-completeness limits imposed in each redshift bin, with higher-$z$ data effectively probing higher-mass galaxies at a given $\beta$.}

\item{Stacking in stellar-mass bins between $2<z\leq 5$ we found the IRX to systematically increase with mass relative to the Calzetti-like IRX-$\beta$ relation. This behavior is consistent with dust-rich systems being characterized by flatter (grayer) attenuation curves, most likely driven by the combined effects of the radiative transfer processes and the increasingly disturbed dust-stars relative morphology. We quantify this relation through the effective slope of the reddening law, $dA_{1600}/d\beta$, where its correlation with the stellar mass can be described by a quadratic relation in $\log(M_\ast/{\rm M_\odot})$, making IRX a bivariate function of $\beta$ and $M_\ast$.}

\item{Expressing IRX directly as a function of $M_\ast$ yields tight correlation between $0.5 < z\leq 5.0$, with IRX rising monotonically with mass. At $z\lesssim 2-3$, however, a clear high-mass turnover was detected, indicating suppressed dust-obscured star formation in massive systems. The corresponding evolution of the $\beta$-$M_\ast$ relation shows a consistent mass-dependent `blueing' at the highest masses at later times, which can be attributed to a transition from a steady cold-gas accretion to phase where the supply of molecular gas is suppressed, limiting dust growth, analogous to the turnover in the star-forming main sequence.}

\item{The relations derived in this work broadly agree with previous stacking-based measurements (e.g. \citealt{McLure_2018, Alvarez_2019,Bouwens_2020}), while providing a consistent framework that deals with the apparent inconsistencies by incorporating the stellar-mass dependence through the slope of the reddening law and more careful SED-based estimation of $\beta$. Differences between our data and other UV-selected samples at low masses can be attributed to stellar mass incompleteness, different methods of dealing with clustering of sources (becoming an increasing issue in low-resolution {\it Herschel} studies), stacking methodology and dust-temperature assumptions.}
    
\end{enumerate}

\begin{acknowledgements}
This research was funded in whole or in part by the National Science Center, Poland (grant no. 2020/39/D/ST9/03078 and 2023/49/B/ST9/00066). For the purpose of Open Access, the author has applied a CC-BY public copyright license to any Author Accepted Manuscript (AAM) version arising from this submission. JSD, DJM acknowledge the support of the Royal Society through the award of a Royal Society University Research Professorship to JSD. KL acknowledges the support of the National Science Centre, Poland, through the PRELUDIUM grant UMO-2023/49/N/ST9/00746. This work is based in part on data products from observations made with ESO Telescopes at the La Silla Paranal Observatory under ESO programme ID 179.A-2005 and ID 179.A-2006 and on data products produced by CALET and the Cambridge Astronomy Survey Unit on behalf of the UltraVISTA and VIDEO consortia. This work is based in part on data obtained as part of the UKIRT Infrared Deep Sky Survey. This work is also based in part on observations made with the {\it Spitzer Space Telescope}, which is operated by the Jet Propulsion Laboratory, California Institute of Technology under NASA contract 1407. This work is based in part on observations obtained with MegaPrime/MegaCam, a joint project of CFHT and CEA/IRFU, at the Canada-France-Hawaii Telescope (CFHT) which is operated by the National Research Council (NRC) of Canada, the Institut National des Science de l'Univers of the Centre National de la Recherche Scientifique (CNRS) of France, and the University of Hawaii.
\end{acknowledgements}

\bibliographystyle{aa_url}
\bibliography{my_papers}

\newpage

\begin{appendix}

\begin{sidewaystable*}
\tiny
\section{Additional tables}\label{sec:ap1}
\caption{IRX values median-stacked in bins of $\beta$ and redshift.}\label{tab:irxb}
\begin{tabular}{ccccccccc}
\hline\hline
& & $-2.25<\beta\leq -1.75$ & $-1.75<\beta\leq -1.25$ & $-1.25<\beta\leq -0.75$ & $-0.75<\beta\leq -0.25$ & $-0.25<\beta\leq 0.25$ & $0.25<\beta\leq 0.75$ & $0.75<\beta\leq 1.25$ \\ 
& $\mathcal{M}_{\rm lim}$ & ${\rm log(IRX)}$ & ${\rm log(IRX)}$ & ${\rm log(IRX)}$ & ${\rm log(IRX)}$ & ${\rm log(IRX)}$ & ${\rm log(IRX)}$ & ${\rm log(IRX)}$ \\ 
& & ${\rm (\beta)}$ & ${\rm (\beta)}$ & ${\rm (\beta)}$ & ${\rm (\beta)}$ & ${\rm (\beta)}$ & ${\rm (\beta)}$ & ${\rm (\beta)}$ \\ 
\hline
\multirow{2}{*}{$2.0<z\leq 2.5$} & \multirow{2}{*}{\phantom{1}9.50} & $-0.35^{+0.30}_{-inf}\phantom{-}$ & $0.15^{+0.19}_{-0.35}$ & $0.74^{+0.07}_{-0.09}$ & $1.36^{+0.04}_{-0.04}$ & $1.91^{+0.03}_{-0.04}$ & $2.30^{+0.03}_{-0.03}$ & $2.71^{+0.02}_{-0.02}$ \\ 
 & & $(-1.82 \pm 0.01)\phantom{-}$ & $(-1.50 \pm 0.01)\phantom{-}$ & $(-1.04 \pm 0.01)\phantom{-}$ & $(-0.55 \pm 0.01)\phantom{-}$ & $(-0.05 \pm 0.10)\phantom{-}$ & ($0.49 \pm 0.01$) & ($0.93 \pm 0.01$) \\ 
\multirow{2}{*}{$2.5<z\leq 3.0$} & \multirow{2}{*}{\phantom{1}9.75} & $0.68^{+0.12}_{-0.17}$ & $0.39^{+0.15}_{-0.23}$ & $0.89^{+0.07}_{-0.08}$ & $1.51^{+0.04}_{-0.04}$ & $2.17^{+0.04}_{-0.04}$ & $2.58^{+0.02}_{-0.02}$ & $3.01^{+0.01}_{-0.01}$ \\ 
 & & $(-1.81 \pm 0.02)\phantom{-}$ & $(-1.47 \pm 0.01)\phantom{-}$ & $(-1.04 \pm 0.01)\phantom{-}$ & $(-0.55 \pm 0.01)\phantom{-}$ & $(-0.03 \pm 0.01)\phantom{-}$ & ($0.45 \pm 0.01$) & ($0.93 \pm 0.10$) \\ 
\multirow{2}{*}{$3.0<z\leq 3.5$} & \multirow{2}{*}{\phantom{1}9.75} & $0.16^{+0.30}_{-inf}$ & $0.60^{+0.12}_{-0.16}$ & $1.01^{+0.07}_{-0.08}$ & $1.54^{+0.04}_{-0.05}$ & $2.27^{+0.02}_{-0.02}$ & $2.61^{+0.02}_{-0.02}$ & $2.94^{+0.01}_{-0.01}$ \\ 
 & & $(-1.85 \pm 0.01)\phantom{-}$ & $(-1.51 \pm 0.01)\phantom{-}$ & $(-1.05 \pm 0.01)\phantom{-}$ & $(-0.57 \pm 0.10)\phantom{-}$ & $(-0.03 \pm 0.10)\phantom{-}$ & ($0.53 \pm 0.01$) & ($1.01 \pm 0.10$) \\ 
\multirow{2}{*}{$3.5<z\leq 4.0$} & \multirow{2}{*}{10.00} & $0.00^{+0.30}_{-inf}$ & $0.92^{+0.10}_{-0.13}$ & $0.99^{+0.11}_{-0.14}$ & $1.70^{+0.05}_{-0.06}$ & $2.32^{+0.02}_{-0.03}$ & $2.50^{+0.03}_{-0.03}$ & $3.00^{+0.02}_{-0.02}$ \\ 
 & & $(-1.85 \pm 0.02)\phantom{-}$ & $(-1.48 \pm 0.01)\phantom{-}$ & $(-1.04 \pm 0.01)\phantom{-}$ & $(-0.56 \pm 0.10)\phantom{-}$ & $(-0.02 \pm 0.10)\phantom{-}$ & ($0.48 \pm 0.14$) & ($0.96 \pm 0.02$) \\ 
\multirow{2}{*}{$4.0<z\leq 5.0$} & \multirow{2}{*}{10.25} & $1.02^{+0.21}_{-0.41}$ & $1.42^{+0.09}_{-0.11}$ & $1.59^{+0.07}_{-0.08}$ & $1.50^{+0.14}_{-0.21}$ & $2.06^{+0.04}_{-0.04}$ & $2.15^{+0.06}_{-0.07}$ & $2.61^{+0.02}_{-0.02}$ \\ 
 & & $(-1.84 \pm 0.10)\phantom{-}$ & $(-1.48 \pm 0.01)\phantom{-}$ & $(-1.04 \pm 0.02)\phantom{-}$ & $(-0.53 \pm 0.10)\phantom{-}$ & ($0.00 \pm 0.10$) & ($0.35 \pm 0.02$) & ($0.96 \pm 0.03$) \\ 
\hline
\end{tabular}
\tablefoot{Data corresponding to color points of Figure\,\ref{fig:irxb}, with $\mathcal{M}_{\rm lim}$ being the mass-completeness limit at each redshift bin, where $\mathcal{M}=\log(M_\ast/{\rm M_\odot})$. Median values of $\beta$ in each bin are listed in the brackets.}

\bigskip\bigskip

\caption{IRX values median-stacked in bins of $\beta$.}\label{tab:irxb1}
\begin{tabular}{ccccccccc}
\hline\hline
& & $-2.25<\beta\leq -1.75$ & $-1.75<\beta\leq -1.25$ & $-1.25<\beta\leq -0.75$ & $-0.75<\beta\leq -0.25$ & $-0.25<\beta\leq 0.25$ & $0.25<\beta\leq 0.75$ & $0.75<\beta\leq 1.25$ \\ 
& $\mathcal{M}_{\rm lim}$ & ${\rm log(IRX)}$ & ${\rm log(IRX)}$ & ${\rm log(IRX)}$ & ${\rm log(IRX)}$ & ${\rm log(IRX)}$ & ${\rm log(IRX)}$ & ${\rm log(IRX)}$ \\ 
& & ${\rm (\beta)}$ & ${\rm (\beta)}$ & ${\rm (\beta)}$ & ${\rm (\beta)}$ & ${\rm (\beta)}$ & ${\rm (\beta)}$ & ${\rm (\beta)}$ \\ 
\hline
\multirow{2}{*}{$2.0<z\leq 5.0$} & \multirow{2}{*}{\rm As\, in\, Table\,\ref{tab:irxb}} & $0.25^{+0.19}_{-0.34}$ & $0.42^{+0.11}_{-0.14}$ & $0.87^{+0.04}_{-0.05}$ & $1.45^{+0.03}_{-0.03}$ & $2.05^{+0.02}_{-0.02}$ & $2.44^{+0.02}_{-0.02}$ & $2.82^{+0.01}_{-0.01}$ \\ 
 & & $(-1.82 \pm 0.01)\phantom{-}$ & $(-1.50 \pm 0.01)\phantom{-}$ & $(-1.04 \pm 0.01)\phantom{-}$ & $(-0.55 \pm 0.03)\phantom{-}$ & $(-0.03 \pm 0.08)\phantom{-}$ & ($0.48 \pm 0.03$) & ($0.94 \pm 0.04$) \\ 
\hline
\end{tabular}
\tablefoot{Same as in Table\,\ref{tab:irxb} but without binning in redshift.}
\end{sidewaystable*}

\begin{sidewaystable*}
\tiny
\caption{IRX-$M_\ast$ stacked values.}\label{tab:irxm}
\begin{tabular}{ccccccccc}
\hline\hline
& $\phantom{1}9.50<\mathcal{M}\leq \phantom{1}9.75$ & $\phantom{1}9.75<\mathcal{M}\leq 10.00$ & $10.00<\mathcal{M}\leq 10.25$ & $10.25<\mathcal{M}\leq 10.50$ & $10.50<\mathcal{M}\leq 10.75$ & $10.75<\mathcal{M}\leq 11.00$ & $11.00<\mathcal{M}\leq 11.25$ & $11.25<\mathcal{M}\leq 11.50$ \\ 
& ${\rm log(IRX)}$ & ${\rm log(IRX)}$ & ${\rm log(IRX)}$ & ${\rm log(IRX)}$ & ${\rm log(IRX)}$ & ${\rm log(IRX)}$ & ${\rm log(IRX)}$ & ${\rm log(IRX)}$ \\ 
& ${\rm (\mathcal{M})}$ & ${\rm (\mathcal{M})}$ & ${\rm (\mathcal{M})}$ & ${\rm (\mathcal{M})}$ & ${\rm (\mathcal{M})}$ & ${\rm (\mathcal{M})}$ & ${\rm (\mathcal{M})}$ & ${\rm (\mathcal{M})}$ \\ 
\hline
\multirow{2}{*}{$0.5<z\leq 1.0$} & $0.04^{+0.30}_{-inf}$ & $0.62^{+0.12}_{-0.17}$ & $0.87^{+0.08}_{-0.10}$ & $1.18^{+0.07}_{-0.08}$ & $1.26^{+0.07}_{-0.08}$ & $1.38^{+0.06}_{-0.08}$ & $1.31^{+0.07}_{-0.09}$ & $1.20^{+0.18}_{-0.31}$ \\ 
 & $\phantom{1}(9.61)$ & $\phantom{1}(9.87)$ & ($10.11$) & ($10.36$) & ($10.61$) & ($10.85$) & ($11.08$) & ($11.32$) \\ 
\multirow{2}{*}{$1.0<z\leq 1.5$} & $0.49^{+0.12}_{-0.17}$ & $0.51^{+0.12}_{-0.16}$ & $0.85^{+0.07}_{-0.08}$ & $1.14^{+0.06}_{-0.07}$ & $1.55^{+0.04}_{-0.05}$ & $1.53^{+0.05}_{-0.06}$ & $1.66^{+0.05}_{-0.06}$ & $1.30^{+0.12}_{-0.17}$ \\ 
 & $\phantom{1}(9.62)$ & $\phantom{1}(9.87)$ & ($10.12$) & ($10.36$) & ($10.61$) & ($10.85$) & ($11.08$) & ($11.33$) \\ 
\multirow{2}{*}{$1.5<z\leq 2.0$} & $0.49^{+0.10}_{-0.14}$ & $0.67^{+0.08}_{-0.10}$ & $0.90^{+0.07}_{-0.08}$ & $1.33^{+0.04}_{-0.04}$ & $1.60^{+0.03}_{-0.04}$ & $1.76^{+0.03}_{-0.04}$ & $1.83^{+0.04}_{-0.04}$ & $1.84^{+0.07}_{-0.08}$ \\ 
 & $\phantom{1}(9.61)$ & $\phantom{1}(9.86)$ & ($10.12$) & ($10.37$) & ($10.61$) & ($10.86$) & ($11.10$) & ($11.32$) \\ 
\multirow{2}{*}{$2.0<z\leq 2.5$} & $0.27^{+0.17}_{-0.27}$ & $0.40^{+0.13}_{-0.18}$ & $0.69^{+0.08}_{-0.10}$ & $1.13^{+0.06}_{-0.08}$ & $1.45^{+0.04}_{-0.05}$ & $1.81^{+0.04}_{-0.05}$ & $2.00^{+0.04}_{-0.04}$ & $2.33^{+0.05}_{-0.06}$ \\ 
 & $\phantom{1}(9.62)$ & $\phantom{1}(9.86)$ & ($10.12$) & ($10.37$) & ($10.61$) & ($10.85$) & ($11.08$) & ($11.32$) \\ 
\multirow{2}{*}{$2.5<z\leq 3.0$} & \multirow{2}{*}{...} & $0.63^{+0.10}_{-0.13}$ & $0.71^{+0.08}_{-0.10}$ & $1.12^{+0.05}_{-0.06}$ & $1.56^{+0.06}_{-0.07}$ & $1.86^{+0.05}_{-0.06}$ & $2.24^{+0.05}_{-0.06}$ & $2.22^{+0.10}_{-0.13}$ \\ 
 & & $\phantom{1}(9.86)$ & ($10.11$) & ($10.36$) & ($10.61$) & ($10.85$) & ($11.09$) & ($11.34$) \\ 
\multirow{2}{*}{$3.0<z\leq 3.5$} & \multirow{2}{*}{...} & $0.66^{+0.10}_{-0.13}$ & $0.75^{+0.08}_{-0.11}$ & $1.02^{+0.07}_{-0.08}$ & $1.53^{+0.06}_{-0.07}$ & $1.92^{+0.07}_{-0.08}$ & $2.51^{+0.06}_{-0.07}$ & $2.94^{+0.09}_{-0.12}$ \\ 
 & & $\phantom{1}(9.87)$ & ($10.11$) & ($10.36$) & ($10.59$) & ($10.86$) & ($11.09$) & ($11.34$) \\ 
\multirow{2}{*}{$3.5<z\leq 4.0$} & \multirow{2}{*}{...} & \multirow{2}{*}{...} & $0.88^{+0.11}_{-0.14}$ & $1.20^{+0.08}_{-0.10}$ & $1.62^{+0.09}_{-0.11}$ & $1.74^{+0.14}_{-0.20}$ & $2.08^{+0.18}_{-0.32}$ & $3.01^{+0.09}_{-0.11}$ \\ 
 & & & ($10.11$) & ($10.35$) & ($10.59$) & ($10.84$) & ($11.08$) & ($11.29$) \\ 
\multirow{2}{*}{$4.0<z\leq 5.0$} & \multirow{2}{*}{...} & \multirow{2}{*}{...} & \multirow{2}{*}{...} & $1.33^{+0.08}_{-0.10}$ & $1.73^{+0.08}_{-0.09}$ & $1.89^{+0.09}_{-0.11}$ & $2.19^{+0.13}_{-0.19}$ & $2.58^{+0.15}_{-0.23}$ \\ 
 & & & & ($10.36$) & ($10.61$) & ($10.86$) & ($11.12$) & ($11.32$) \\ 
\hline
\end{tabular}
\tablefoot{Data corresponding to color points of Figure\,\ref{fig:irxm1} and black circles of Figure\,\ref{fig:irxm}, with $\mathcal{M}=\log(M_\ast/{\rm M_\odot})$. Median values of $\mathcal{M}$ in each bin are listed in the brackets.}
\end{sidewaystable*}

\end{appendix}

\end{document}